\documentclass[format=acmsmall, review=false, screen=true]{acmart}
\usepackage{booktabs} 
\pdfoutput=1 

\acmJournal{TKDD}
\acmVolume{0}
\acmNumber{1}
\acmArticle{00}
\acmMonth{9}

\setcopyright{acmcopyright}

\setcitestyle{numbers}

\acmDOI{0000001.0000001}

\usepackage{multibib}
 \newcites{latex}{Data and Code References}
 
\usepackage{longtable}
\usepackage[export]{adjustbox}

\usepackage{booktabs}
\usepackage[utf8]{inputenc}
\usepackage[font={small,bf},labelfont=bf,format=plain]{caption}
\usepackage{subcaption}
\usepackage{textcomp}
\usepackage{amsmath}
\usepackage{enumitem}
\usepackage{algorithm}
\usepackage{algpseudocode}
\algtext*{EndProcedure}
\usepackage{xcolor,colortbl}
\usepackage{footnote}
\usepackage{url}
\usepackage{tabularx}
\usepackage{multicol}
\usepackage{multirow}
\usepackage{pifont}
\usepackage{bm}
\usepackage{listings}
\usepackage{wrapfig}
\usepackage{relsize}
\usepackage{mathtools}
\usepackage{bbm}
\usepackage{blindtext}
\usepackage{bm}
\usepackage{balance}
\usepackage{mdwlist}
\setlist[enumerate]{leftmargin=*}
\setlist[itemize]{leftmargin=*}

\usepackage{wasysym}
\usepackage{pifont}

\newtheorem{observation}{Observation}

\newcommand\mytop{\rule{0pt}{2ex}}

\newcommand{\embedding}{\mathbf{E}}
\newcommand{\similarity}{\mathbf{S}}
\newcommand{\adj}{\mathbf{A}}
\definecolor{mygreen}{RGB}{34,139,34}

\newcommand{\matA}{\mathbf{A}}
\newcommand{\matS}{\mathbf{S}}
\newcommand{\matP}{\mathbf{P}}

\begin{abstract}
    While most network embedding techniques model the proximity between nodes in a network, recently there has been significant interest in \textit{structural embeddings} that are based on node \textit{equivalences}, a notion rooted in sociology: 
    equivalences or positions are collections of nodes that have similar roles---i.e., similar functions, ties or interactions with nodes in other positions---irrespective of their distance or reachability in the network.
    Unlike the proximity-based methods that are rigorously evaluated in the literature, the evaluation of structural embeddings is less mature. 
    It relies on small synthetic 
    or real networks  with labels that are not perfectly defined, and its connection to sociological equivalences has hitherto been vague and tenuous. 
    With new node embedding methods being developed at a breakneck pace, \textit{proper evaluation and systematic characterization of existing approaches will be essential to progress.}
    
    To fill in this gap, we set out to understand \textit{what} types of equivalences structural embeddings capture. 
    We are the first to  contribute rigorous intrinsic and extrinsic evaluation methodology for structural embeddings, 
    along with carefully-designed, diverse datasets of varying sizes. 
     We observe a number of different evaluation variables that can lead to different results (e.g., choice of similarity measure, classifier, label definitions).  
    We find that degree distributions within nodes' local neighborhoods can lead to simple yet effective baselines in their own right and guide the future development of structural embedding. 
    We hope that our findings can influence the design of further node embedding methods and also pave the way for more comprehensive and fair evaluation of structural embedding methods. 
    
\end{abstract}

\begin{document}

\title{Towards Understanding and Evaluating Structural Node Embeddings}

\author{Junchen Jin}
\affiliation{
  \institution{University of Michigan}
  \city{Ann Arbor}
  \state{MI}
  \country{USA}
}
\email{kinmark@umich.edu}

\author{Mark Heimann}
\affiliation{
  \institution{University of Michigan}
  \city{Ann Arbor}
  \state{MI}
  \country{USA}
}
\email{mheimann@umich.edu}

\author{Di Jin}
\affiliation{
  \institution{University of Michigan}
  \city{Ann Arbor}
  \state{MI}
  \country{USA}
}
\email{dijin@umich.edu}

\author{Danai Koutra}
\affiliation{
  \institution{University of Michigan}
  \city{Ann Arbor}
  \state{MI}
  \country{USA}
}
\email{dkoutra@umich.edu}

\date{}

\maketitle

\section{Introduction}
\label{sec:intro}

Node embeddings capture similarity between nodes in a multi-dimensional space:  
the closer two nodes are embedded, the more similar they are in the network.
Two broad categories of node similarity  are prevalent in the representation learning literature:
(i)~proximity, which intuitively embeds similarly nodes that belong to communities or cohesive groups~\cite{deepwalk,line}; and
(ii)~equivalence or structural similarity, which aims to similarly embed nodes that have similar patterns of relations with other nodes 
irrespective of their exact location in the graph~\cite{wasserman_faust_1994,rossi2019community}.

In this work, we focus on \textit{structural node embeddings}, which preserve structural similarity. 
Unlike proximity-preserving embedding methods that model first- or second-order proximity~\cite{line}, or sample neighborhood context via random walks~\cite{gemsurvey,deepwalk,node2vec,rossi2019community}, 
structural embeddings are inspired from the notions of roles and positions in sociology. 
A \textit{position} or \textit{equivalence class} describes a collection of individuals with similar \textit{roles}, i.e., similar functions, ties or interactions with individuals in other positions~\cite{wasserman_faust_1994}. 
Depending on the type of equivalence (e.g., automorphic, regular---cf. Section~\ref{sec:equivalences}), different positions and roles arise that enable  
both multi-network tasks (e.g., network alignment and classification~\cite{xnetmf,rossi2019community}, transfer learning~\cite{rolx}) 
and single-network tasks, including structural role classification, 
anomaly detection, and identity resolution~\cite{node2bits}.
To capture the notion of roles in the network, structural embeddings  are typically based on feature-based matrix factorization~\cite{rolx,xnetmf} or random walks~\cite{struc2vec}, graphlets~\cite{role2vec}, or more recently LSTMs~\cite{drne}.

While proximity-based methods are evaluated rigorously on a set of well-understood tasks using established datasets, the evaluation of structural embeddings is less mature. 
It relies mostly on limited experiments on a barbell graph,  
or structural node classification / clustering of \textit{small} real datasets (mainly air-traffic networks) with node labels whose definitions are contrived.
It also lacks rigorous connections to the types of equivalence from which role discovery in networks stems. 

Our goal is to contribute toward the systematic evaluation of unsupervised feature representations of nodes. In natural language processing, evaluation of unsupervised word representations has long been recognized as an important area of study.  Prominent works have as their  objective the standardization of evaluation of word embeddings~\cite{schnabel2015evaluation}.  Other works have pointed out additional evaluation methods and challenges to the point where a multi-year workshop has arisen dedicated to the evaluation of word embeddings\footnote{\url{https://repeval2019.github.io/}}, and the field of word embedding evaluation now warrants a survey~\cite{bakarov2018survey}. Node embedding, being a comparatively newer area of study, is only now starting to see similar growth, and the recent works that have focused on intrinsic~\cite{dalmia2018towards} or extrinsic evaluation~\cite{goyal2019benchmarks,gurukar2019network} of node embeddings focus only on proximity-preserving embeddings.  Interest in structural embeddings, has been growing, however, and a recent survey distinguishes them from proximity-preserving embeddings~\cite{rossi2019community}.   A standardized analysis of structural embedding methods is essential to ensure that the problem area indeed continues to see forward progress.

Toward this end, we provide 
{\bf a novel, comprehensive evaluation methodology for systematic analysis of structural embedding methods with respect to the sociological theories of equivalence}. Our main contributions are: 

\begin{itemize}
    \item \textbf{Evaluation Methodology.} This is the first paper to introduce a variety of evaluation methods for \textit{unsupervised structural node embeddings} (Section~\ref{sec:prelim}). These are based on:  
    (i)~intrinsic measures related to equivalence definitions (Section~\ref{sec:equivalences}), which help us decouple the effectiveness of methods from classifiers in downstream tasks, and 
    (ii)~extrinsic measures that characterize their performance in the context of high-value tasks (Section~\ref{sec:tasks}), for which we rethink the ground truth used in prior work.

\item \textbf{Appropriate Datasets.} We introduce new benchmark datasets, and ways to obtain ground truth roles (Section~\ref{sec:data}). We hope that these datasets will change the way structural embeddings are evaluated.

\item \textbf{In-depth Empirical Analysis.} 
Our empirical analysis of 11 methods (Section~\ref{sec:methods}) on 35 real and synthetic datasets and a variety of tasks (Section~\ref{sec:properties}-\ref{sec:emb-equivalences}) shows that different methods win based on different metrics, label definitions, downstream machine learning models, or embedding similarity functions (e.g., cosine vs.\ Euclidean). {This analysis highlights that there is no one optimal structural embedding.}
Moreover, we evaluate the extent to which sociological equivalences are captured by different structural embedding methods (Section~\ref{sec:emb-equivalences}).  Also, besides merely comparing the performance of different methods on downstream tasks, we further analyze their performance at a finer granularity to understand for \textit{which types of nodes} current methods perform best (Section~\ref{sec:equiv-labels-extrinsic}).    
    
\item \textbf{New Design Insights.} We find that degree distribution in nodes' local neighborhoods is effective as a feature representation in its own right as well as the building block for some of the most successful embedding methods.  This can influence the design of future structural embedding methods and/or serve as a standalone baseline for structural embedding tasks.  

\end{itemize}

\vspace{0.05cm}
For reproducibility and wider adoption of structural node embeddings by the community, we make our code available in a Python package, which includes the implementations of 11 structural embedding approaches: \url{https://github.com/GemsLab/StrucEmbedding-GraphLibrary} along with our extensive experimental benchmarks.  Its modular design is meant to be easy to use as well as to extend by adding new datasets or embedding methods, and we envision it facilitating standardized experimental evaluation in subsequent research works.  

Next, we present our methodology and research goals.

\vspace{-0.2cm}
\section{Methodology}
\label{sec:prelim}
In this section, we first introduce node embedding and describe in more detail several methods that we empirically analyze in this work (Section~\ref{sec:methods}).  To understand better what structural node embeddings learn, we turn to concepts introduced in other academic disciplines to analyze the structural roles of nodes: role equivalences in mathematical sociology (Section~\ref{sec:equivalences}), as well as statistics developed by network scientists to measure node connectivity and centrality (Section~\ref{sec:statistics}).  We then present the tasks for which we evaluate node embeddings (Section~\ref{sec:tasks}), and finally the goals of our research study (Section~\ref{sec:goals}).  

\subsection{Node Embedding Methods}
\label{sec:methods}

Node embedding is a function mapping nodes $V$ in a network $G$ to $d$-dimensional feature vectors $\mathbf{x}\in\mathbb{R}^d$ such that ``similar'' vertices have similar feature vectors, {based on some similarity measure}.  
In this work, we propose a thorough evaluation methodology for \textit{structural} node embeddings. 
These methods assume two nodes are similar if they have similar structural roles (defined in Section~\ref{sec:equivalences}) \textit{regardless} of their proximity in the network, and `hybrid' approaches 
(i.e., that capture both proximity and structural similarity to some extent).  We refer the reader to ~\cite{rossi2019community} for  more information on this distinction. 

In our analysis, we consider a large number of existing \textit{unsupervised} node embedding methods, predominantly structural embeddings but with a few proximity-preserving node embedding methods as well for contrast. We also introduce three simple variants of degree distributions as baselines. Since we propose intrinsic evaluation, which is not dependent on a downstream task, we do not include supervised methods (e.g., graph neural networks) in our empirical study.
We introduce the methods that we consider below by category. We also provide their code versions and detailed descriptions of their hyperparameter configurations in Appendix~\ref{app:embed-param}.

\vspace{0.15cm}
\noindent \textbf{Proximity-based (or hybrid) embeddings.} In our analysis, we consider two embedding methods that are primarily proximity-based. 
\textbf{(1)~node2vec}~\cite{node2vec} uses the skip-gram architecture~\cite{word2vec} to learn an embedding for each node that preserves its similarity to other nodes in its context, sampled with biased random walks. 
\textbf{(2)~LINE}~\cite{line} optimizes an embedding objective that maximizes the probability of the first and second-order proximities in the network (direct edges between any two nodes and mutual neighbors that any two nodes share, respectively).  
Proximity methods are the topic of numerous surveys~\cite{gemsurvey,rossi2019community}, and we refer the interested reader to those. 

\vspace{0.15cm}
\noindent \textbf{Structural embeddings.} We also evaluate {eight} structural embedding approaches: 
\textbf{(3)~struc2vec}~\cite{struc2vec} uses the same skip-gram architecture, but samples context with random walks performed over an auxiliary multilayer graph capturing structural similarity (mainly \textit{degree}) of nodes' neighborhoods at several hop distances.  
\textbf{(4)~GraphWave} \cite{graphwave} computes the heat kernel matrix for a graph and embeds each node by sampling the empirical characteristic function of the distribution of heat it sends to other nodes.
\textbf{(5)~xNetMF}~\cite{xnetmf} draws on the connection between the skip-gram architecture matrix factorization~\cite{levy2014neural} to find node embeddings that implicitly factorize a structural similarity matrix, 
defined by comparing the distribution of node degrees in $k$-hop neighborhoods. 
{Subsequently, \textbf{(6) SEGK}~\cite{segk} factorizes a structural similarity matrix using graph kernels to compare the nodes' $k$-hop neighborhoods.} 
\textbf{(7)~role2vec}~\cite{role2vec} applies the skip-gram model to a corpus sampled using \emph{attributed} random walks which record the structural type of each node. The method learns the same embedding for nodes of each structural type, which enhances space efficiency. 
{\textbf{(8)~RiWalk}~\cite{riwalk} also uses the skip-gram model, but learns an embedding for each node based on the structural types of nodes in its context.}  
\textbf{(9)~DRNE}~\cite{drne} contends that feature propagation is similar to the recursive definition of regular equivalence, and uses an LSTM to learn node embeddings by aggregating the features of their neighbors sorted sequentially by degree.  
\textbf{(10)~MultiLENS}~\cite{multilens}, similar to xNetMF, derives embeddings based on matrix factorization that captures the distribution of structural features in nodes' local neighborhoods.  While we focus on unsupervised methods in this paper and thus do not consider common semi-supervised graph neural network models~\cite{kipf2017semi, graphsage}~\footnote{The unsupervised objective proposed to train the graph neural network GraphSAGE~\cite{graphsage} models node proximity rather than structural roles, more akin to a method like node2vec~\cite{node2vec}.}, it has been noted that MultiLENS performs local feature aggregation akin to that of a graph neural network~\cite{multilens}.

\vspace{0.15cm}
\noindent \textbf{Other structural methods.} In addition to these ten `hybrid' and structural methods, 
we also construct \textbf{(11)} variants of \textbf{degree distributions} over different neighborhoods, which can be seen as simple, yet strong, baselines for embedding nodes. 
We represent each node with the degree distribution of its $k$-hop neighbors---i.e, a histogram of dimension $d_{max}$, the maximum node degree in each dataset, in which the $i$-th entry counts the number of neighbors that are $k$ hops away with degree $i$.
We refer to the {11$^{th}$} family of structural approaches that we consider as \textbf{\texttt{degree}} that is simply the node's degree, and \textbf{\texttt{degree1}} and \textbf{\texttt{degree2}} that are histograms based on 1- and 2-hop neighborhoods.

\subsection{Equivalence in Social Science}
\label{sec:equivalences}

Structural embeddings are related to the notions of \textit{social roles} or \textit{positions}, which are central in sociology for understanding how the society or groups are organized.
\textit{Role} refers to the patterns of relations between individuals, or the ways in which individuals relate to each other. 
\textit{Position} or \textit{equivalence class} describes a collection of individuals with similar activity, ties or interactions with individuals in other positions~\cite{wasserman_faust_1994}.
The formal definitions of these terms are based on network methods, which led to their wide adoption in social network analysis.
In network analysis, (structural) roles of nodes include centers of stars, peripheral nodes, bridge nodes, members of cliques, and more~\cite{rolx}.

There are different types of equivalence, each of which is based on an equivalence relation that defines a partition of a node-set to mutually exclusive and exhaustive equivalence classes such that the nodes that are equivalent are assigned to the same class.
Among the various types of equivalence, we focus on three main types: structural, automorphic, and regular equivalence. 

\begin{figure}[t]
    \centering
    \includegraphics[width=.75\linewidth]{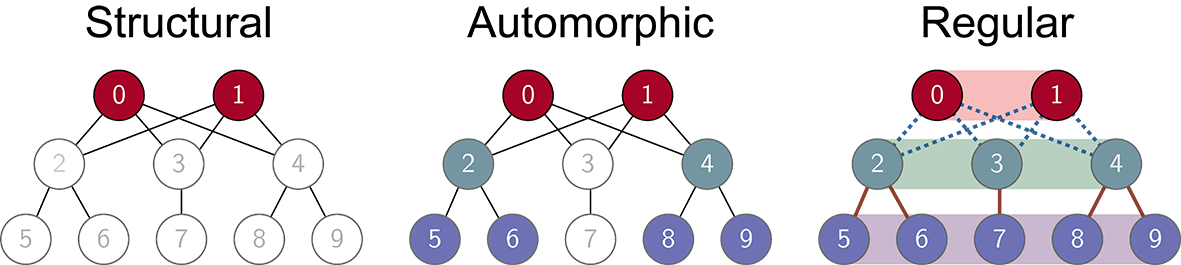}
    \vspace{-0.26cm}
    \caption{Different types of equivalence. Nodes filled with the same color belong to the same 
    equivalent roles.}
    \label{fig:equivalences}
    \vspace{-0.3cm}
\end{figure}

\vspace{0.1cm}
 \textbf{Structural equivalence}~\cite{Lorrain} is the simplest and most restrictive notion of equivalence:
\begin{definition} 
\vspace{-0.12cm}
Two nodes are {\em structurally equivalent} iff they have identical connections with identical nodes.
\end{definition}

\vspace{-0.12cm}
\noindent For example, in Figure~\ref{fig:equivalences} nodes $0$ and $1$ are structurally equivalent. Structural equivalence is rarely seen in real-world networks, and it is very strict form of structural similarity that is closely related to proximity: \textit{two structurally equivalent nodes are at most two hops away from each other}~\cite{wasserman_faust_1994,RossiA15}. We confirm empirically that proximity-preserving embedding methods best capture this in 
Section~\ref{sec:emb-equivalences}.

\vspace{0.1cm}
 \textbf{Automorphic equivalence}~\cite{BorgattiE92_equiv} was proposed to relax the notion of structural equivalence. Intuitively, two automorphically equivalent nodes are identical with respect to \textit{all} graph theoretic properties (e.g., in-/out-degree, centralities) and may differ only in terms of their labels. Examples include the nodes in each node-set \{$0$, $1$\}, \{$2$, $4$\}, and \{$5$, $6$, $8$, $9$\} of Figure~\ref{fig:equivalences}. More formally:
\begin{definition} 
\vspace{-0.12cm}
Two nodes are {\em automorphically equivalent} iff there is an automorphism (i.e., an isomorphism in the \emph{same} graph) that maps one node to the other.
\end{definition}

\vspace{-0.12cm}
Although automorphic equivalence is less restricted than structural equivalence (and also a superset of structural equivalence), its \textit{exact} format is still expected to be rare in real networks.

\vspace{0.1cm}
 \textbf{Regular equivalence}~\cite{BorgattiE92_equiv} is among the most interesting and prevalent types of equivalence in real networks, and it is defined as follows: 
\begin{definition}
\vspace{-0.12cm}
Two nodes are {\em regularly equivalent} if they relate in the same way to \emph{equivalent} nodes. 
\label{dfn:regular}
\end{definition}

\vspace{-0.12cm}
\noindent This definition is more meaningful in multi-relational networks (e.g., heterogeneous graphs), but it also applies to networks with a single relation. 
For example, similarly colored nodes in Figure~\ref{fig:equivalences} correspond to regularly equivalent classes---e.g., nodes \{2,3,4\} are regularly equivalent because they connect to nodes of the `red' and `purple' roles, although they do not have the same degree (and, thus, it is more relaxed notion than automorphic equivalence).

\subsection{Network Statistics}
\label{sec:statistics}
Network scienctists have proposed several statistics that capture the structural properties of a node in a network. 
To gain a better understanding of what properties structural node embeddings capture, we consider four popular graph statistics: 
\begin{itemize}
\item \textbf{Degree} is the most direct measure of a node's connectivity.  In the unweighted, undirected networks that are usually considered for structural node embedding, a node's degree is the number of connections (equivalently, the number of incident edges or neighboring nodes) it has.  

\item \textbf{PageRank} is a node importance score, recursively calculated for each node based on the importance scores of its neighbors.  

\item \textbf{Betweenness centrality} of a node describes how many shortest paths in a graph between \emph{other} pairs of nodes go through that node.  

\item \textbf{Clustering coefficient} describes for each node, how many of its neighbors are also connected. 
\end{itemize}

The first three statistics are all measures of centrality: nodes with high degree, pagerank, or betweenness centrality are generally considered well connected in the graph and play a prominent structural role.  They range from local to global: a node's degree can be computed purely off its local connections.  A node's pagerank score is computed from its neighbors, but due to the recursive definition of the pagerank scores, nodes may end up aggregating information globally from across the graph (although distant nodes influence its pagerank score much less than closer neighbors).  Betweenness centrality is explicitly calculated based on global information from all pairs of other nodes.  Clustering coefficient, on the other hand, is less a measure of centrality and instead describes how densely connected a node's neighborhood is.  Again, this is a local score that can be computed individually for each node based only on its immediate neighborhood.

\subsection{Tasks}
\label{sec:tasks}
\label{sec:tasks}
Node embeddings may be used for a variety of downstream tasks.  To evaluate the utility of various methods, we compare them based on several families of tasks which we discuss here.  

\subsubsection{Single-Network Tasks}
Structural node embeddings are often used to predict the labels of nodes, when these are thought to correspond to a node's structural role in a network~\cite{struc2vec}.  The problem of \textbf{node classification} can be modeled as a supervised machine learning problem that can be solved with any off-the-shelf downstream machine learning classifier~\cite{scikit-learn} applied to the embeddings of the nodes.  A related unsupervised task is \textbf{node clustering}~\cite{graphwave}, which can also be solved with standard machine learning methods applied to the features of the nodes obtained via embeddings.   
Link prediction seeks to infer whether two unconnnected nodes should share an edge.  It is a common task for node embeddings~\cite{node2vec,BelthBK20}; however, the fundamental insight needed for link prediction is the proximity of the nodes (whether or not they should share an edge and be in close proximity).  This task is thus more suitable for proximity-preserving node embeddings, and we do not study it further in this work.  

\subsubsection{Multi-Network Tasks}
Structural node embeddings have also been shown to be useful for tasks defined over multiple networks, as structural roles can be compared across networks.  A task that exemplifies node comparison across networks is \textbf{network alignment}, where the objective is to find correspondences betwen nodes in different networks.  Nodes may be matched based on similarity of their structural node embeddings~\cite{xnetmf}.  The structural embeddings for all nodes in a network can also be aggregated into a single feature vector for the entire network, which may be used for graph-level tasks like \textbf{graph classification}~\cite{rgm}.  Thus, any structural node embedding method can be used for graph alignment or classification by employing the recently proposed embedding-based methods for these tasks~\cite{xnetmf,rgm}.  

\subsection{Research Goals}
\label{sec:goals}
In this work, our goal is to help the research community understand and evaluate structural node embeddings more thoroughly.  We contribute to the understanding and evaluation of existing structural node embedding methods, but also with our analysis pave the way for better understanding and evaluation of methods that are subsequently developed.  

\subsubsection{Understanding} Fundamentally, we want to learn what aspects of a ``structural role'' do node embedding capture.  Here, we turn to concepts of role equivalence developed in mathematical sociology (Section~\ref{sec:equivalences}), as well as network-scientific statistics (Section~\ref{sec:properties}), to see how well each embedding method captures these properties. Existing works claim to capture different types of equivalences without detailed justifications; we aim to clarify previous misconceptions and incorrect claims through careful empirical analysis.

\subsubsection{Evaluation} We propose new methods for intrinsic as well as extrinsic evaluation of structural node embeddings.  Intrinsic evaluation directly evaluates the geometry of the node embedding space, independent of any downstream task or method (e.g., a machine learning classifier).  The goal is to see how similarities between nodes in the embedding space correlate to similarities defined by a ground-truth task or by the sociological and network scientific concepts we introduce.  Extrinsic evaluation, on the other hand, analyzes the performance of a downstream task using the node embeddings.  We cast our objectives for understanding as an extrinsic evaluation, by using machine learning to predict role equivalences or network statistics from the node embeddings.  We also consider the single- and multi-network tasks discussed in Section~\ref{sec:tasks}.

To study structural embedding methods meaningfully, we need datasets that highlight what they are able to capture.  Toward this end, we collect real datasets on which the data mining task of interest (e.g., node labels for classification) relates to the structural roels of each node.  We also design an extensive generation of synthetic datasets, going beyond the simpler constructions of previous works~\cite{struc2vec, graphwave} specifically to illustrate clear role equivalences (Section~\ref{sec:equivalences}).  

While we empirically study a large majority of existing structural embedding methods, the primary purpose of our evaluation is not to choose a ``winner'' from existing structural embedding methods. We see that various methods have their own strengths and weaknesses, and indeed our contributions are forward-looking with the goal of positively influencing the development of future structural embedding methods.  To be most constructive for future developments in structural embedding methods:
\begin{itemize}
    \item We identify factors unrelated to the node embeddings that may influence the ranking embeddings.  In Section~\ref{sec:real-exp}, we show that on the same graph, different structural embedding methods may appear to be better or worse due to a variety of factors, like the performance metric, distance metric used to compare node embeddings, downstream machine learning model, or definition of node labels.  Future works should be mindful of these to avoid reporting apparent gains that are due to some factor other than the quality of the embeddings.  
    \item We highlight successful (and unsuccessful) design choices for different tasks.  We design a simple set of baselines, local degree histograms, that are based on design choices that appear to perform well in many of the tasks we consider.  Future works can not only compare against these baselines, but also use the ideas they incorporate to develop more effective methodology.  
\end{itemize}

\section{Data and Ground Truth Roles}
\label{sec:data}

\begin{table}[b!]
\vspace{-0.3cm}
\caption{Real Datasets: Single-Network Tasks}
\label{tab:real}
\vspace{-0.2cm}
{\small 
    \begin{tabular}{lrrl}
    \toprule
        \textbf{Dataset} & \textbf{\# Nodes} & \textbf{\# Edges} & \textbf{Labels} \\ \midrule
        \textbf{BlogCatalog}~\cite{node2vec} &10,312 &333,983 & centralities  \\
        \textbf{Facebook}~\cite{node2vec}& 4,039 &88,234 & equivalences (Section~\ref{sec:equivalences}) \\
        \textbf{ICEWS}~\cite{icews} & 1,255 & 1,414 & military vs media entities \\
        \textbf{Email-300} &318 &752 & professional roles \\
        \textbf{Email-2K} & 2,414 &11,995 & professional roles \\
        \textbf{PPI}~\cite{graphsage} & 56,944 &818,786 & protein cellular functions \\
        \textbf{BR air-traffic}~\cite{struc2vec} &  131 & 1,038 & \# landings \& take-off, equival. (Section~\ref{sec:equivalences})\\
        \textbf{EU air-traffic}~\cite{struc2vec}  &  399 & 5,995 & \# landings \& take-off, equival. (Section~\ref{sec:equivalences})\\
        \textbf{US air-traffic}~\cite{struc2vec}  &  1,190 & 13,599 & \# passengers, equivalences (Section~\ref{sec:equivalences})\\
        \textbf{DD6}~\cite{BorgwardtK05} &4,152 &20,640 & amino acid properties\\
        \bottomrule
    \end{tabular}
    }
\end{table}

\begin{figure}[b!]
    \centering
    \begin{subfigure}[b]{0.52\columnwidth}
        \centering
         \includegraphics[width=\linewidth,trim={0 0 25cm 0},clip]{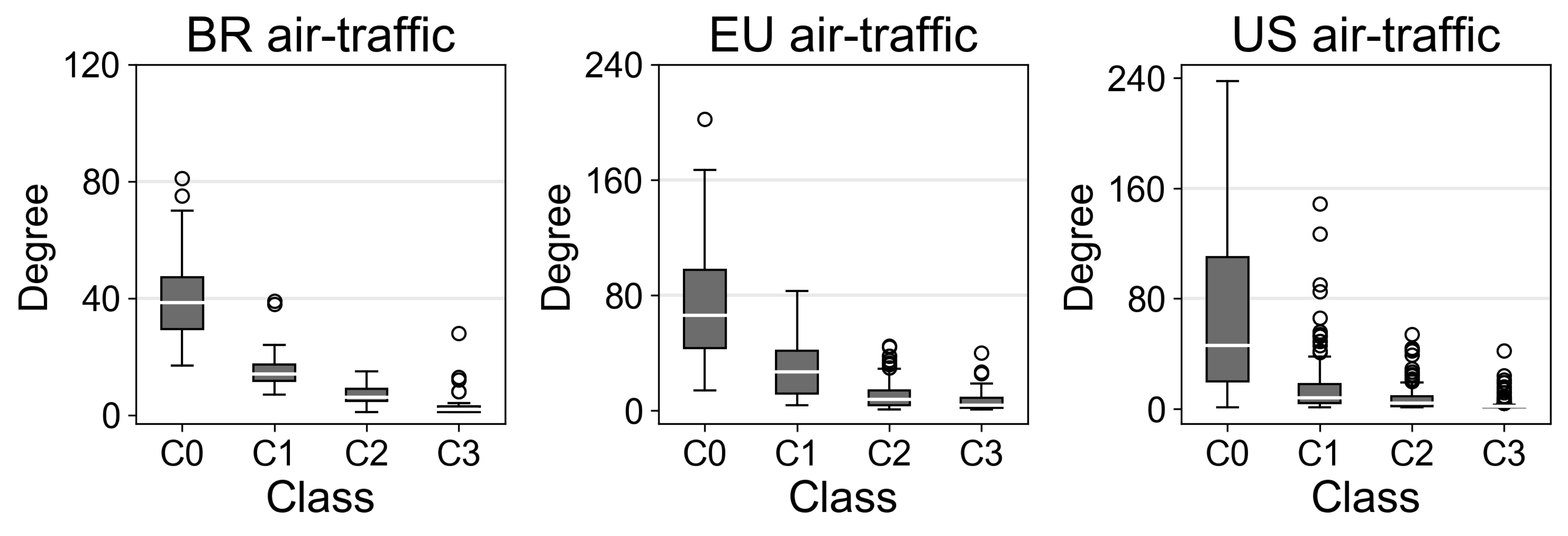}
    \vspace{-0.5cm}
    \caption{
    Strong connection between the node degree and the class labels in the BR and EU air-traffic data.}
    \label{fig:class_vs_degree}
    \end{subfigure}
    \hfill
    \begin{subfigure}[b]{0.38\columnwidth}
        \centering
        \includegraphics[width=.6\linewidth]{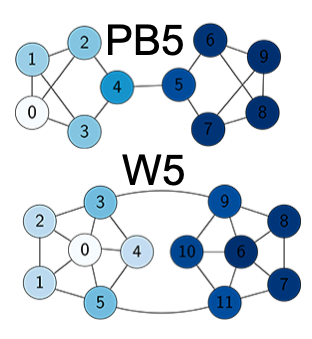}
        \vspace{-0.2cm}
        \caption{Vertex similarity~\cite{VertexSimilarity} is related to proximity. 
        Lighter color represents higher similarity to node $0$ (in white).}
        \label{fig:vertex_similarity}
    \end{subfigure}
    \vspace{-0.3cm}
    \caption{Limitations of some node labeling methods.}
\end{figure}

\begin{figure*}[t]
    \centering
    \includegraphics[width=.9\linewidth]{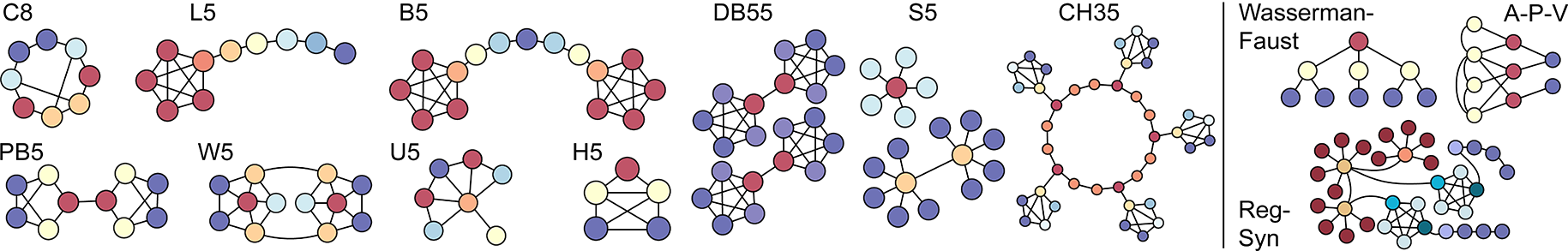}
    \vspace{-0.3cm}
    \caption{Per synthetic base graph, nodes with the same color are automorphically equivalent on the left \& regularly equivalent on the right.}
    \label{fig:synthetic_panel}
    \vspace{-0.2cm}
\end{figure*}

To gain 
insights into the type of information that is encoded in structural embeddings, we consider several real datasets (Table~\ref{tab:real}), and introduce synthetic data (Figure~\ref{fig:synthetic_panel}, Table~\ref{tab:enlarged_network_s_a}), the structure of which we can control and understand better than that of real networks. 

Our datasets feature unweighted, undirected graphs without node or edge attributes, as this is the traditional paradigm studied in structural node embedding~\cite{struc2vec, graphwave} and is the simplest common denominator of graph structure that all the methods in our study can handle.  We note that several of the structural embedding methods we consider, including xNetMF~\cite{xnetmf} and GraphWave~\cite{graphwave} can be extended to weighted and directed graphs~\cite{ember} or even signed networks~\cite{heimann2020structural}.  While we do not study such graph contexts here, we expect that as more structural node embedding methods are developed specifically for them, the standards we set for evaluation can also be extended.  

\subsection{Real Network Data: Single-Network Tasks}
\label{subsec:real-data}
\subsubsection{Limitations of existing datasets}
\label{sec:limitations}
The most commonly used real datasets for evaluating the quality of structural embeddings are \textbf{air-traffic networks} from \cite{struc2vec}, which capture the existence of commercial flights (edges) between airports (nodes) and are thus undirected and unweighted~\cite{struc2vec}. Their node labels 
are defined based on either the number of landings and take-offs, or the number of passengers passed by each airport in a given time period: four labels are obtained by splitting the data into quartiles.  Although the balanced classes simplify the evaluation, this arbitrary labeling has two drawbacks: (1) it is not clear that splitting the data into four quartiles reflects a real-world phenomenon; 
and (2) to a large extent, the labels simply capture degree information (Figure~\ref{fig:class_vs_degree}). 

To experiment with the effect of different node labelings to the performance, we also construct an alternative set of node labels constructed by splitting the airport-related statistics (number of landings and take-offs, or passengers) into logarithmic bins (Figure~\ref{fig:results-by-label}).  This results in imbalanced classes but produces a distribution of ``roles'' following the well-known power-law distribution.

More recent work~\cite{drne} also used a Jazz collaboration network and BlogCatalog, creating labels using the \textit{vertex similarity} measure~\cite{VertexSimilarity} as ground truth for regular equivalence. However, as we show in Figure~\ref{fig:vertex_similarity}, the vertex similarity captures {\em distances} between nodes rather than similarity in their structural properties, and thus is not an appropriate measure for regular equivalence.

\subsubsection{New datasets for structural embeddings.} Besides the existing datasets used in prior works on structural embeddings, we also consider large real-world datasets (Table~\ref{tab:real}), where we can define the node labels based on 
the different definitions of equivalence (Section~\ref{sec:equivalences},\ref{sec:emb-equivalences}).   We use the \textbf{BlogCatalog} and \textbf{Facebook} networks from \cite{node2vec}, which are both social network datasets containing various structural roles.

Real-world data mining tasks are often defined in terms of external node labels, so to this end we propose the use of additional datasets where this information may be better predicted by structural rather than proximity-preserving node features.   The first is a knowledge graph of the relationships among socio-political actors from the Integrated Crisis Early Warning System (\textbf{ICEWS})~\cite{icews}; it is constructed from events on October 4, 2018 that are automatically extracted from news articles.  We group the entity types into broad categories, and our task is to distinguish between ``media'' entities and ``military'' entities.  We expect that these will have distinct structural roles from each other.  
Another real dataset we use is the \textbf{PPI network} from \cite{graphsage}, a multi-network dataset which is claimed to have node labels corresponding to structural roles rather than communities. 
Finally, we use a network called \textbf{DD6}, one of the larger networks from the D\&D dataset commonly used to benchmark graph classification~\cite{BorgwardtK05}.  This dataset is a protein structure and its nodes, which represent amino acids, have labels representing various properties of the amino acid~\cite{BorgwardtK05}.  These labels exhibit very low homophily and are known to be challenging for proximity-based node representation learning methods~\cite{lee19-motif-attention}. We also use two proprietary email communication networks, \textbf{Email-300} and \textbf{Email-2K}, for the users in which we have professional roles (e.g., CEO, manager) that are known to be related to regular equivalence~\cite{wasserman_faust_1994}.

\subsubsection{Ground-truth Node Equivalences or Roles}
\label{subsec:UCINET}
For our intrinsic evaluation, instead of arbitrarily defining roles in networks, we leverage existing (exact or approximate) algorithms that aim to identify equivalence classes. 
Given the adjacency matrix $\matA$ of a graph, these approaches produce a pairwise node similarity matrix $\matS$ based on their respective equivalence definitions.
For \textit{structural equivalence}, CONCOR~\cite{CONCOR} creates a similarity matrix with entries $s_{ij}=s_{ji}$ corresponding to the Pearson correlation between nodes $i$ and $j$ (i.e., the correlation of their respective rows, $\matA_{i,:}$ and $\matA_{j,:}$).
For \textit{automorphic equivalence}, MAXSIM~\cite{EVERETT198877} first creates a matrix of geodesic proximities from the adjacency matrix $\matA$, and then creates $\matS$ by comparing the node distributions of geodesic proximities pairwise.
For \textit{regular equivalence}, CATREGE~\cite{CATREGE} searches for matches in successive node neighborhoods,
and encodes in $\matS$ the iteration in which two nodes were separated into different groups or classes.

CONCOR also produces a partition 
that we use as the \textit{ground-truth} equivalence classes 
(i.e., groups of nodes with similar roles).   To obtain the ground truth for 
MAXSIM and CATREGE, 
we apply hierarchical clustering on $\matS$ (with default settings). 

\subsection{Real Network Data: Multi-Network Tasks}
While structural node embeddings are often used for single-network tasks such as node classification and clustering, recent works have used them for multi-network tasks such as network alignment~\cite{xnetmf} and classification~\cite{rgm}.  In Section~\ref{sec:multinetwork}, we comprehensively evaluate a large number of structural embedding methods within the embedding-based frameworks proposed to solve these downstream tasks.  Here we describe the standard benchmark datasets we use for each tasks.  

For network classification, we use three well-known and publicly available~\cite{DD6_repo} graph classification benchmark datasets, PTC-MR, IMDB-M, and NCI1 . These correspond to small, medium, and large graph classification datasets as used in recent work~\cite{rgm}.  IMDB-M is a social network dataset where the graphs represent actor collaboration networks, and in other two the networks represent small molecules.  The molecular datasets also have node labels, which to fairly compare all embedding methods we do not use in the embeddings, but which can be used by a downstream graph classification method.  We give detailed descriptions of the datasets in Table~\ref{tab:datasets-graphclassification}.

\begin{table}[t!]
\centering
{ \small
    \caption{Graph classification datasets~\cite{DD6_repo}. We give the total number of nodes/edges across all graphs per dataset.} 
    \label{tab:datasets-graphclassification}
    \begin{tabular}{l r@{\hspace{2pt}}  r@{\hspace{2pt}} @{\hspace{2pt}}r @{\hspace{2pt}}r @{\hspace{2pt}}c @{\hspace{3pt}}l}
    \toprule
       \textbf{Name} & \textbf{Nodes} & \textbf{Edges} &  \textbf{Graphs} &  \textbf{Classes} & \textbf{Node labels} & \textbf{Domain} \\
    \midrule
         PTC-MR & 4\,916 & 5\,053 & 344 & 2 & Y & bioinformatics \\

         IMDB-M & 19\,502 & 98\,910 & 1\,500 & 3 & N & collaboration \\ 
         
         NCI1 & 122\,765 & 132\,753 & 4\,110 & 2 & Y & bioinformatics \\ 
         
    \bottomrule
    \end{tabular}
}
\end{table}

For network alignment, we use two datasets from ~\cite{xnetmf}, which again represent social and biological phenomena.  We describe the process of constructing a network alignment scenario with known ground-truth correspondences between nodes, which is commonly used in the network alignment literature, in Section~\ref{sec:multinetwork}.

\begin{table}[t!]
\centering
\caption{Graph alignment datasets.} 
\label{tab:datasets-alignment}
{\small
\begin{tabular}{l r@{\hspace{1.2em}}  r@{\hspace{4pt}} @{\hspace{5pt}}r l}
\toprule
   \textbf{Name} & \textbf{Nodes} & \textbf{Edges} &  \textbf{Description}  \\
\midrule
     Arenas Email \cite{koblenz} &  1\,133      & 5\,451    & communication network \\
     
     PPI \cite{ppi} & 3\,890     & 76\,584    &  PPI network (Human) \\

\bottomrule
\end{tabular}
}
\end{table}

\subsection{Synthetic Network Data}
\label{subsec:synthetic-data}

We also evaluate structural embedding techniques on a variety of synthetically-generated networks---beyond just the commonly-used barbell graph---, as shown in Figure~\ref{fig:synthetic_panel} (left). 

We define two sets of roles per node, based on \textit{structural} and \textit{automorphic}---using the methods CONCOR and MAXSIM (Section~\ref{subsec:UCINET}), respectively. We also enlarge the small synthetic graphs to enable further extrinsic evaluation (Table~\ref{tab:enlarged_network_s_a}).
For \textit{regular} equivalence, since nodes should be assigned to different classes according to their roles, we generate the synthetic graphs accordingly (Figure~\ref{fig:synthetic_panel}, right). Similarly, we  enlarge the synthetic graphs by adding more nodes with different roles and connecting them following the rules in the base case (Table~\ref{tab:enlarged_network_s_a}). For all the synthetic graphs generated for the regular equivalence evaluation, the edge type is indicated by the pre-defined roles of the end-points (e.g., hub vs.\ clique node). The output of CATREGE (Section~\ref{subsec:UCINET}) generates the \textit{same} role assignment as the pre-defined roles. 

\renewcommand{\arraystretch}{1.3}
\begin{table}[t]
\caption{
Enlarged synthetic graphs}
\label{tab:enlarged_network_s_a}
\centering
{\small 
    \begin{tabular}{l@{\hskip 0.1in}c@{\hskip 0.1in}p{9cm}}
\toprule 
  \textbf{Large Graph} &\textbf{Base} &\textbf{Generation}\\\hline
  \textbf{H10\_S\_L} &\textbf{H5} & 10 H5 on a circle with 2 circular nodes between each connecting circular node with house's side.\\
  \textbf{H10\_T\_L} &\textbf{H5} & 10 H5  on a circle with 2 circular nodes between each connecting circular node with  house's roof.\\
  \textbf{Barbell L-A} &\textbf{B5} & Connecting the out-most nodes on the chain of B5 into a circle.\\
  \textbf{Barbell L-B} &\textbf{B5} & Connecting the out-most nodes on the chain of B5 into a circle. Additional 5-clique at each connector.\\
  \textbf{Ferris Wheel} &\textbf{C8} & Enlarged version of C8 with similar perturbation.\\
  \textbf{City of Stars} &\textbf{S5} & 10 normal stars and 5 binary stars as in S5\\
  \textbf{PB-L} &\textbf{PB5} & 10 half-sided PB5 connected to each node of a 10-node circular graph. All the node degrees are 3.\\  \hline
  \textbf{Conference} &\textbf{A-P-V} & Mimicking the real-world scenario, we simulate 80 papers with 4$\sim$6 collaborators out of the 120 authors, and assign them to one of the 30 venues.\\ 
  \textbf{Reg-Syn-L} &\textbf{Reg-Syn} & Based on the connection rules in Reg-Syn, we connect 9 stars, 7 cliques and 7 chains of different sizes.\\
  \textbf{Knitting Wheel} &\textbf{B5} & 10 different sized cliques connected onto a circle with three circular nodes apart each connection.\\
    \bottomrule 
    \end{tabular}
}
\end{table}
\renewcommand{\arraystretch}{1}

\section{Embeddings and Structural Properties}
\label{sec:properties}
Many of the existing structural embedding methods (Section~\ref{sec:methods}) leverage node degree information in various ways.
While it is expected that embeddings are \textit{related} to the node degrees, it is not well-understood to \textit{which extent} they capture the degree or other structural information (e.g., centralities). 
In this section, we seek to gain insights into this via correlation and predictive analysis.  While such an analysis will not completely characterize the information captured in structural embeddings, it can help us understand which ones are comparatively \emph{interpretable} in the sense that they encode common network metrics used to characterize a node's structural role.  

\subsubsection*{Methodology}
First, to see if similarly embedded nodes have similar structural properties, we perform the following analysis: 
    {\bf (1)}~For each node $v$ in graph $G$, we calculate a property of interest $p_i(v)$. We consider four properties: degree, PageRank (with damping parameter $\alpha=0.85$~\cite{ilprints422}), clustering coefficient, and betweenness centrality.
    {\bf (2)}~We identify $v$'s $k$-nearest neighbors ($k$-NN) in the embedding space $\mathbb{R}^d$ using cosine or Euclidean distance, and compute the average value for each structural property, $\overline{p_{i,kNN}(v)}$. 
    {\bf (3)}~Per property $p_i$, we calculate the Pearson correlation between the structural property of a node and its $k$-NN across all nodes.

\begin{figure*}[t!]
    \centering
    \begin{subfigure}[b]{0.48\textwidth}
        \centering
        \includegraphics[width=.95\linewidth]{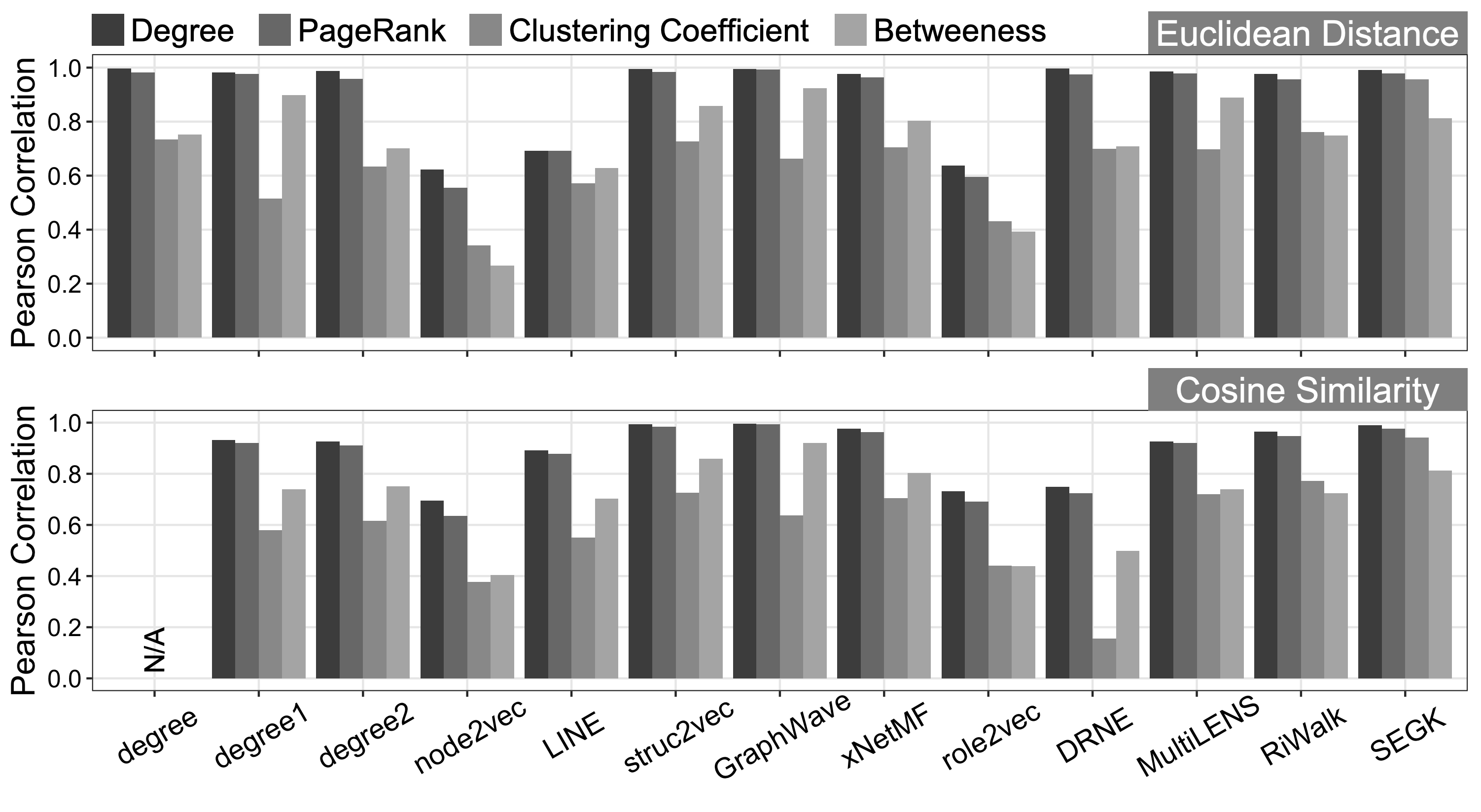}
    \caption{EU air-traffic: correlation between structural properties of a node and the structural properties of its $5$-NN in the embedding space.}
    \label{fig:correlation-blogcatalog}
    \end{subfigure}
    \hfill
    \begin{subfigure}[b]{0.48\textwidth}
        \centering
        \includegraphics[width=\linewidth]{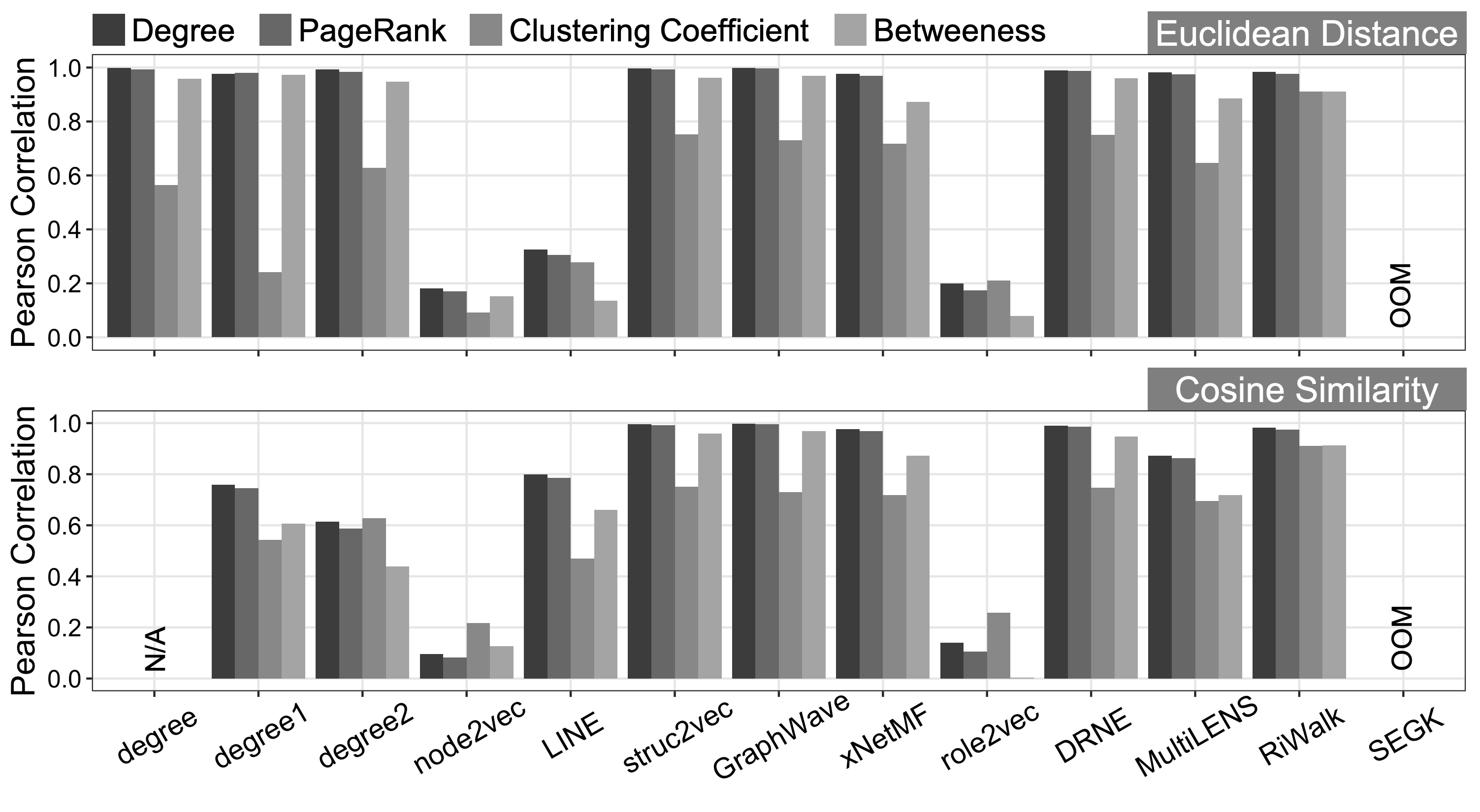}
        \caption{BlogCatalog: correlation between structural properties of a node and the structural properties of its $5$-NN in the embedding space.}
        \label{fig:correlation-eu}
    \end{subfigure}
    \caption{Correlation of embeddings with structural properties: Generally, structural methods---except role2vec---do well in preserving the node structural properties in the embedding space $\mathbb{R}^{d}$. Degree and PageRank are better captured than betweenness and clustering coefficient. As expected, proximity-based embedding methods don't perform well. Differences are observed between Euclidean distance and cosine similarity.}
    \label{fig:correlations}
\end{figure*}

Second, to better understand the extent to which degree is encoded in the structural embeddings, we also perform a predictive task. 
Given a subset of nodes with their structural embeddings and degrees, we apply $k$-NN regression and compute the error between the predicted and original degree for the remaining nodes. We report the mean RMSE across 5 folds, using one fold for \textit{training} and four folds for \textit{testing}.

\begin{figure}[t!]
    \centering
    \vspace{-0.35cm}
    \includegraphics[width=.9\linewidth]{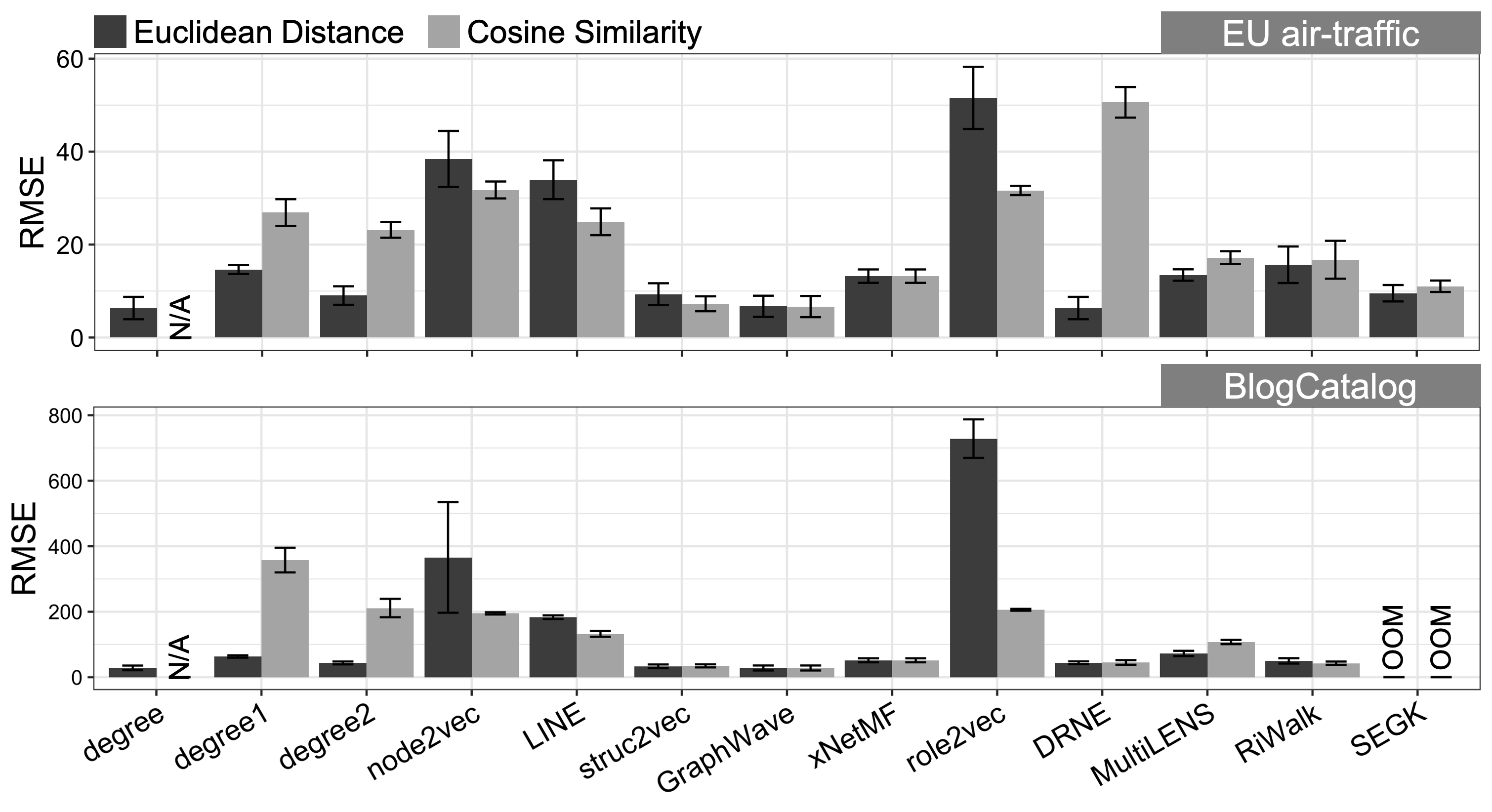}
    \vspace{-0.5cm}
    \caption{RMSE of predicting the node degree from the structural embeddings for two datasets: BlogCatalog (top, max degree=3,992) and EU Air-traffic network (bottom, max degree=202). Error bar shows standard deviation on 5 fold CV with one fold as training and four folds as testing. Performance on the predictive task aligns with the correlation task. Choice of distance metric influences the performance of some methods significantly (e.g., DRNE, role2vec).}
    \label{fig:rmse}
    \vspace{-0.3cm}
\end{figure}

\vspace{-0.15cm}
\subsubsection*{Results}  Since this task is based on the intrinsic properties of the embeddings, and not the node labels, theoretically we can use any dataset here. We report results on the prevailing BlogCatalog and EU air-traffic network datasets. The results are consistent on other data (real and synthetic). The cosine distance is not defined between pure scalars so we leave the result for the degree variant with cosine as N/A (Not Applicable) in all the results.

Based on Figure~\ref{fig:correlations}, for most structural embedding methods, except role2vec, closely embedded nodes have similar degree/PageRank centralities. 
These embeddings also contain information about betweenness and clustering coefficient, but less so.  On the BlogCatalog dataset, RiWalk preserves betweenness and clustering coefficient almost as well as degree and pagerank; most other methods have a significant drop in at least one of the two former metrics, as does RiWalk on the EU Air-traffic dataset. Proximity-based embedding methods such as node2vec and LINE do not encode structural properties well. 

\begin{observation}
Current structural node embeddings capture node importance measures such as degree and pagerank well, but discern less clearly the density of connectivity as given by betweenness and clustering coefficient.
\end{observation}

The results for correlation in Figure~\ref{fig:correlation-eu} and~\ref{fig:correlation-blogcatalog} differ for Euclidean distance and cosine similarity, especially for proximity methods. A further discussion on the usage of similarity measurement is in Section~\ref{sec:conclusions}. 

Similar patterns can be observed from the RMSE in the predictive task (Figure~\ref{fig:rmse}) 
with $5$-NN regression. The maximum node degree for BlogCatalog is 3,992 and EU air-traffic network is 202. With only 20\% of the node's degrees as training, struc2vec, GraphWave, xNetMF and MultiLENS can perform well on the predictive task.

\section{Embeddings and Equivalences}
\label{sec:emb-equivalences}
In the literature, there are various claims about the types of equivalence that embedding methods capture, some of which are imprecise. 
We investigate this by designing experiments for both intrinsic and extrinsic evaluation. 
Our \textbf{intrinsic evaluation} aims to evaluate the quality of embeddings in the context of different types of equivalences \textit{directly},  decoupled from a downstream task. 
Here, \textit{ground-truth labels} are defined by the equivalence methods (Section~\ref{sec:equivalences}, \ref{subsec:UCINET}).
Our \textbf{extrinsic evaluation} relies on classification and clustering, both of which are typically used to evaluate embeddings.

\subsection{Intrinsic Evaluation}
\label{sec:synthetic_intrinsic}

The intrinsic evaluation of structural embeddings seeks to characterize the agreement between the similarities of nodes defined by the different types of equivalence and the node similarities in the embedding space $\mathbb{R}^d$.

\subsubsection{Methodology.} 
\label{sec:synthetic_intrinsic_methodology}

Given a similarity matrix $\matS$ based on a notion of role equivalence (\ref{subsec:UCINET}), for each node we calculate the Kendall rank correlation coefficient between its embedding similarity (based on Euclidean distance or cosine similarity\footnote{It is not defined for a scalar (e.g., \texttt{degree}), in which case we list "N/A" in Figs.~\ref{fig:rank-avg}-\ref{fig:rank-ex}.}) and its structural similarity to all other nodes given by $\similarity$. 

For structural and automorphic equivalence, we perform analysis on a total of 16 synthetic networks (Figure~\ref{fig:synthetic_panel} left plus the enlarged datasets in the top section of Table~\ref{tab:enlarged_network_s_a}, CH35 excluded as near-duplication of Small Town-S) and 4 real networks (three air-traffic networks + Facebook). One exception is that for structural equivalence, CONCOR encounters an error for City of Stars, for which we skipped evaluation. For regular equivalence, we perform analysis on a total of 5 synthetic datasets (Figure~\ref{fig:synthetic_panel} right plus the enlarged datsets in the bottom section of Table~\ref{tab:enlarged_network_s_a}, A-P-V excluded as duplication of Conference). None of our real networks can be used with CATREGE to compute regular equivalence for an intrinsic evaluation, as the algorithm requires relationship types and the implementation handles up to 255 nodes. 
For each type of equivalence, we report the average and the standard deviation of the Kendall rank correlation coefficient across different subsets of our datasets.

\subsubsection{Results.} 
\label{sec:intrinsic-results}

\begin{figure}[hp]
    \centering
    \begin{subfigure}[b]{\columnwidth}
            \centering
            \includegraphics[width=.96\linewidth,frame]{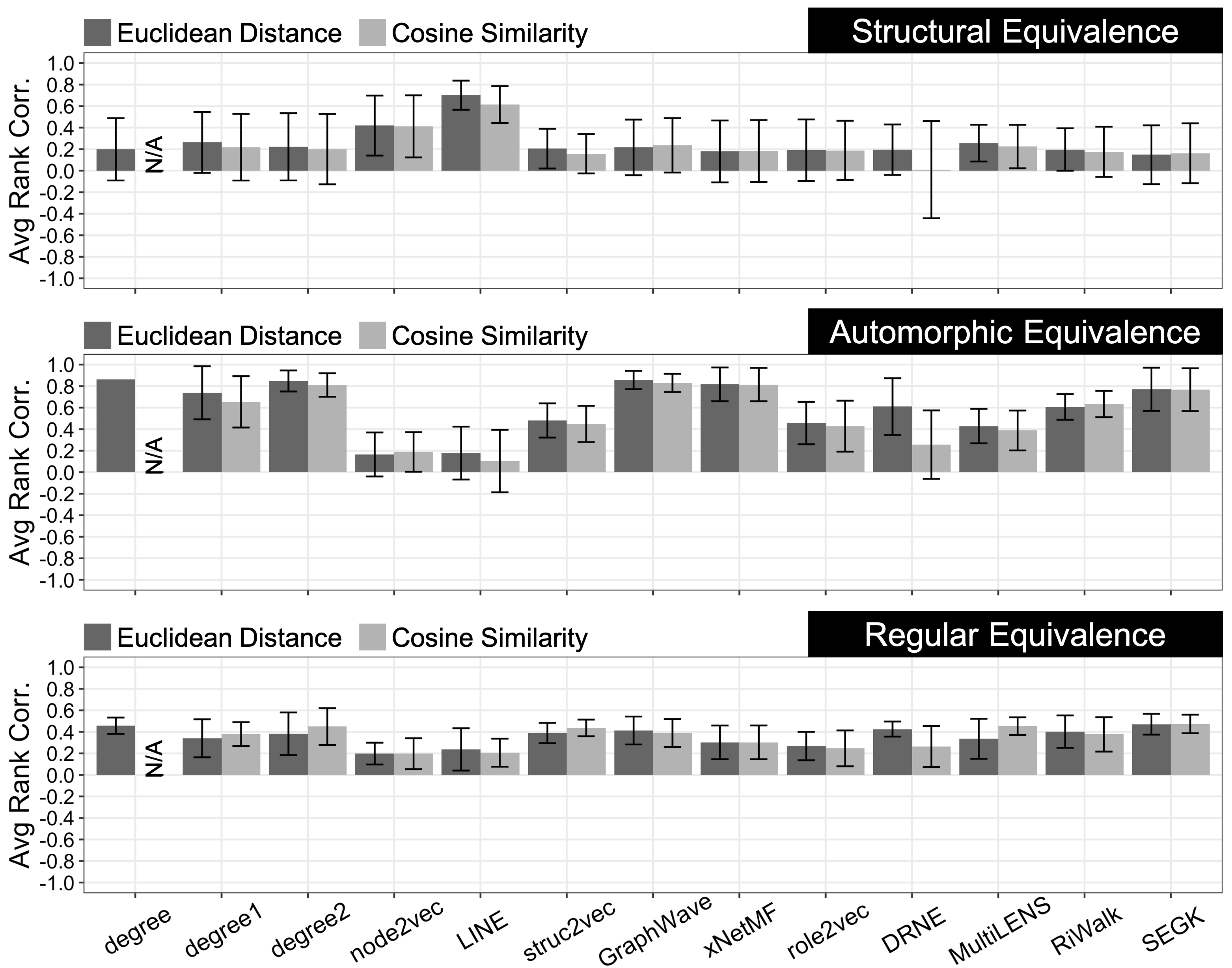}
            \label{fig:rank-synthetic}
        \vspace{-0.2cm}
        \caption{Synthetic data}
        \end{subfigure}
    
        \begin{subfigure}[b]{\columnwidth}
            \centering
            \includegraphics[width=.96\linewidth,frame]{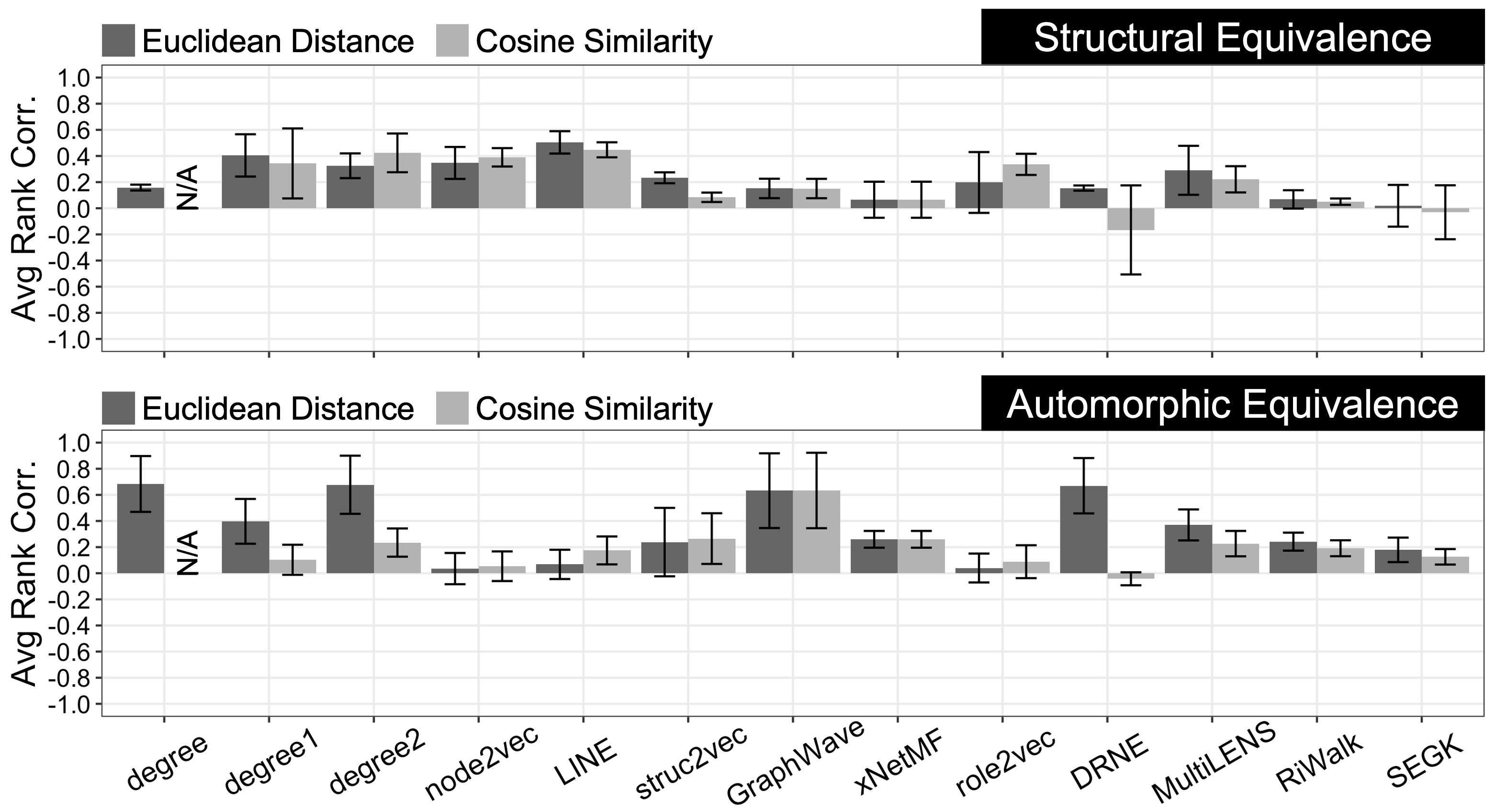}
            \vspace{-0.2cm}
            \caption{Real data (no ground truth for regular equivalence)}
            \label{fig:rank-real}
        \end{subfigure}
        \vspace{-0.65cm}
        \caption{Summarized view of intrinsic evaluation: Average correlation (and stdev) between node embeddings and different types of equivalences across synthetic data (top) and real data (bottom). 
        Structural embeddings tend to capture automorphic and regular equivalence, while primarily proximity embeddings capture structural equivalence. The choice of distance affects the results. } 
\label{fig:rank-avg}
\vspace{-0.55cm}
\end{figure}

Figure~\ref{fig:rank-avg} gives a summarized view of our intrinsic evaluation. It shows, per embedding method, the rank correlation and its standard deviation averaged over all the corresponding datasets.
 LINE and node2vec rank top in our intrinsic evaluation for \textbf{structural equivalence}. This is expected, as despite its name, structural equivalence is actually by definition best captured by proximity-based embedding methods~\cite{wasserman_faust_1994,RossiA15}.  It is defined between two nodes in terms of how many neighbors they share: two nodes are structurally equivalent if they are connected to the exact same nodes.   
Structural equivalence as defined in mathematical sociology is distinct from the structural similarity that role-based node embeddings try to capture.

\begin{figure}[hp]
   \centering
    \begin{subfigure}[b]{\columnwidth}
        \centering
        \includegraphics[width=.7\linewidth]{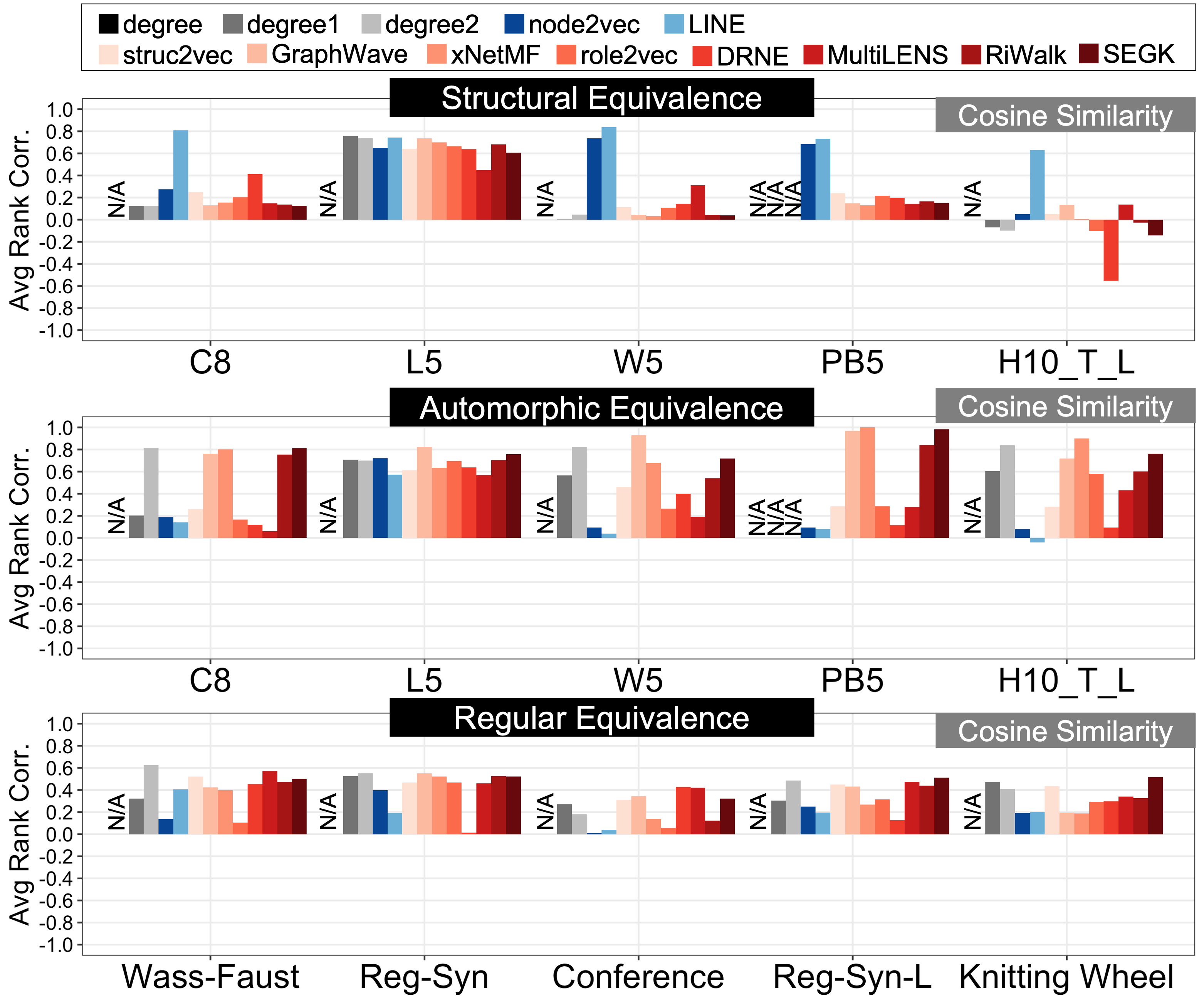}
    \caption{Synthetic data (only cosine similarity shown for brevity) }
    \label{fig:rank-synthetic-ex}
    \end{subfigure}

    \begin{subfigure}[b]{\columnwidth}
        \centering
        \includegraphics[width=.7\linewidth]{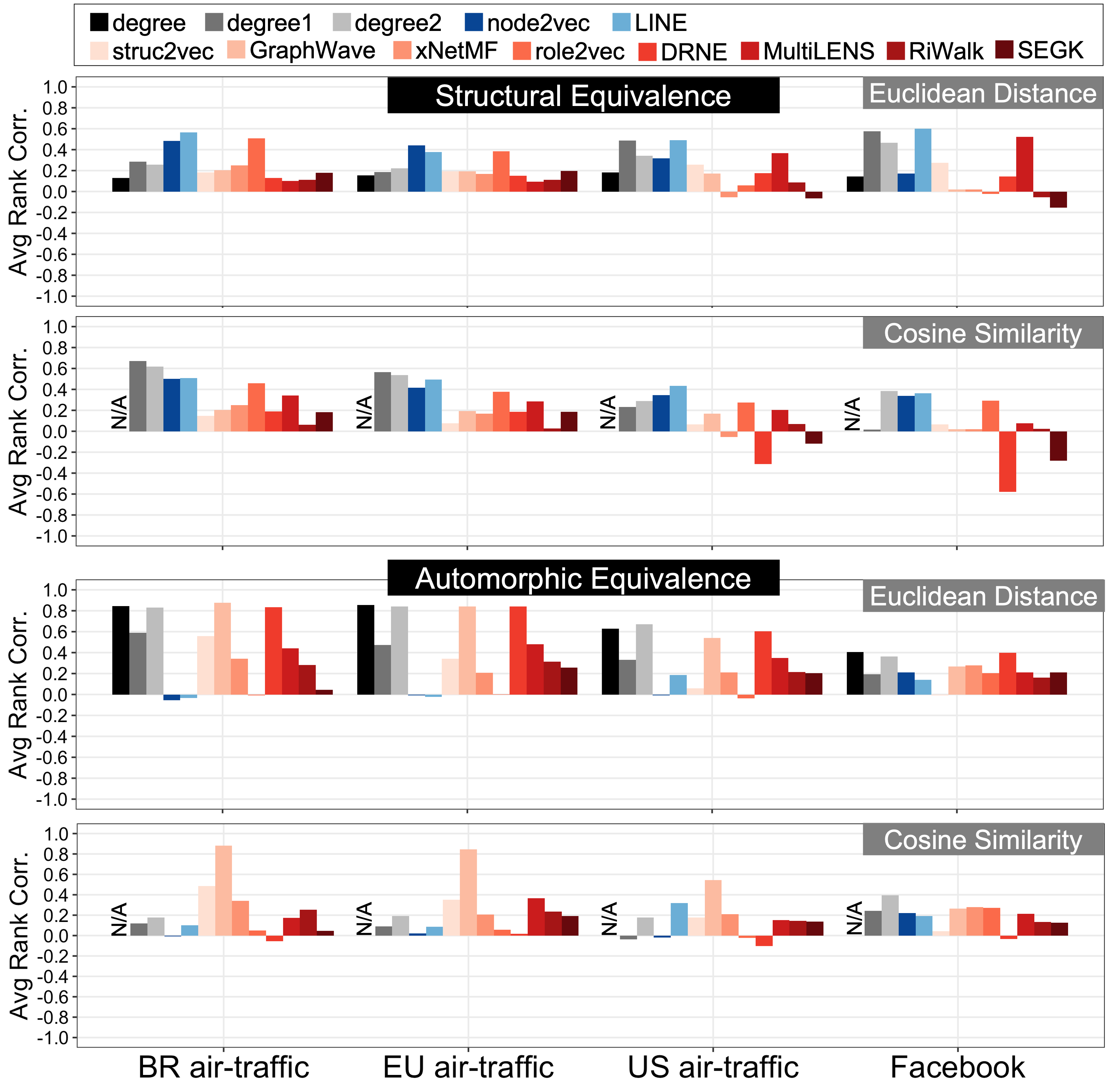}
        \caption{Real data (no ground truth for regular equivalence)} 
        \label{fig:rank-real-ex}
    \end{subfigure}
    \caption{[Best viewed in color]
    Detailed view of intrinsic evaluation: correlation with different types of equivalence for specific synthetic (top) and real (bottom) datasets. 
    Performance of embedding methods varies across different datasets and distance choices.} 
\label{fig:rank-ex}
\end{figure}

\begin{observation}
Structural equivalence depends on node proximity and in fact cannot be captured well by structural embeddings, but automorphic equivalence does not depend on node proximity and may be better captured by structural embeddings than by proximity-preserving embeddings.
\end{observation}

On the other hand, structural embedding methods such as GraphWave, xNetMF and SEGK, as well as \texttt{degree2}, work well in terms of \textbf{automorphic equivalence}, while the proximity-based methods, like LINE and node2vec do not.  This is also expected, as automorphically similar nodes need \textit{not} be in close proximity in the graph. 
We conjecture that the difference of role2vec on the synthetic datasets and real world datasets might result from the difference in degree distribution and network structure between the synthetic and real datasets. 

Similarly, the proximity-based node2vec and LINE struggle to capture \textbf{regular equivalence}, which among structural embedding methods is generally best captured by \texttt{degree}, DRNE, and GraphWave based on Euclidean distance, and \texttt{degree2}, MultiLENS, and struc2vec based on cosine similarity.  The strong performance of degree distribution features in the intrinsic evaluation using automorphic and regular equivalence is noteworthy. 

\begin{observation}
Node degree, generalized to include the distribution in its $k$-hop neighborhood, may indeed be a good indicator of the structural position or role of the node in the network. 
\end{observation}

\enlargethispage{\baselineskip}

In Figure~\ref{fig:rank-ex}, we look deeper into these results on a per-dataset basis.  While trends are largely similar, some datasets are worth noting individually. For example, we see that the base ``L5'' has a distinctive ``lollipop'' shape, where equivalent nodes (in the head) and comparatively near-equivalent nodes (in the stem) are also in close proximity.  As a result, proximity-preserving and structural embeddings do comparably well at capturing both structural and automorphic equivalence. We see larger gaps on the remaining synthetic datasets. On real datasets, GraphWave and DRNE capture extremely high automorphic equivalence on the air-traffic datasets, but the difference between them and the other methods disappears on Facebook, a social network dataset.  

\begin{observation}
None of the structural embedding methods are optimized to capture sociological concepts of role equivalence.
\end{observation}
  
Although we find that structural embedding methods do capture sociological role equivalence to some extent incidentally, it depends on how well the equivalences correspond in any given dataset with the types of similarities each embedding is optimized to preserve (the choice of distance, Euclidean or cosine, has significant impact for some methods, especially in the real data.)

\subsection{Extrinsic Evaluation}
\label{sec:equiv-labels-extrinsic}

Next we seek to evaluate the structural embeddings \textit{extrinsically}  by defining \textit{equivalence-specific} node labels. 

\subsubsection{Methodology.} 
As described in Section~\ref{sec:synthetic_intrinsic_methodology}, we consider the equivalence-specific similarity matrix
$\matS$ and the network embeddings $\embedding$. 
To obtain the ground-truth \textit{equivalence classes} (i.e., node labels), we perform hierarchical clustering on $\matS$ for MAXSIM and CATREGE, and use the CONCOR partitioning output directly (Section~\ref{subsec:UCINET}). Again, for the synthetic datasets used for automorphic equivalence evaluation, we manually define the \textit{exact} automorphically equivalent classes (instead of using MAXSIM, which is an approximation). With the classes generated or pre-defined, we perform classification and clustering for extrinsic evaluation.

\vspace{0.1cm}
\noindent \textbf{Classification Setup.}
For each dataset, we use 5-fold cross validation to get the average performance and standard deviation. A multinomial logistic regression with $\ell_2$ penalty, $C=1.0$ is trained to perform multi-class classification. {The other parameters are set as default from the scikit-learn package \cite{scikit-learn}}. We use Accuracy and $\text{Macro-F}_{\text{1}}$ to evaluate the performance of classification.

\vspace{0.1cm}
\noindent \textbf{Clustering Setup.}
Per dataset, we use $k$-means to cluster the embeddings $\embedding$, setting $k$ to be the number of ground truth clusters. To mitigate the effects of algorithmic instability, we run $k$-means for 1,000 times with different centroid seeds and use the best output in terms of the inertia criterion. We use Normalized Mutual Information (NMI) and purity scores to evaluate clustering performance.

\begin{figure*}[t!]
    \centering
    \begin{subfigure}[b]{0.47\textwidth}
        \centering
        \includegraphics[width=\linewidth]{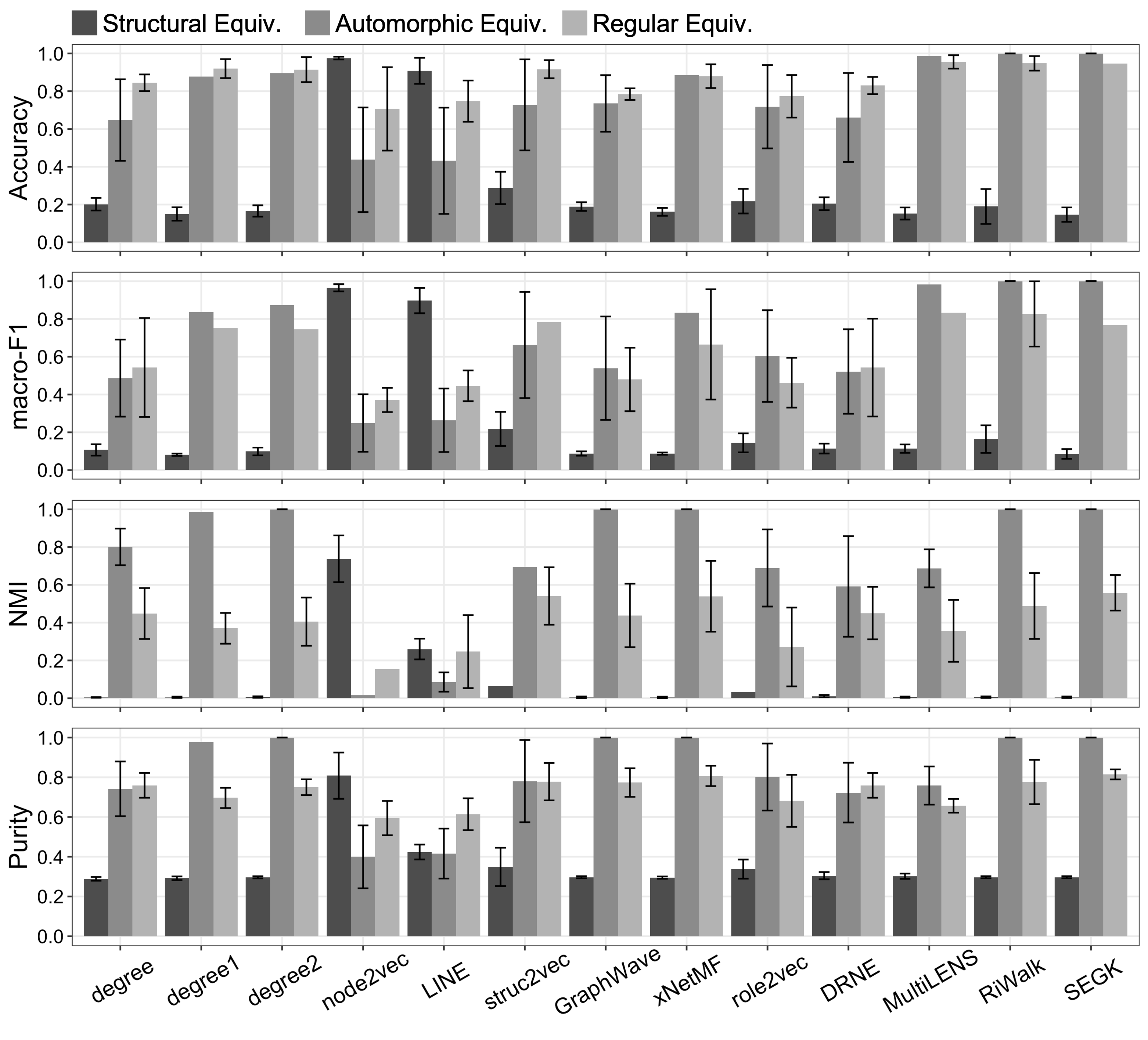}
        \label{fig:downstream-synthetic}
    \vspace{-0.65cm}
    \caption{Synthetic data}
    \end{subfigure}
    \hfill
    \begin{subfigure}[b]{0.47\textwidth}
        \centering
        \includegraphics[width=\linewidth]{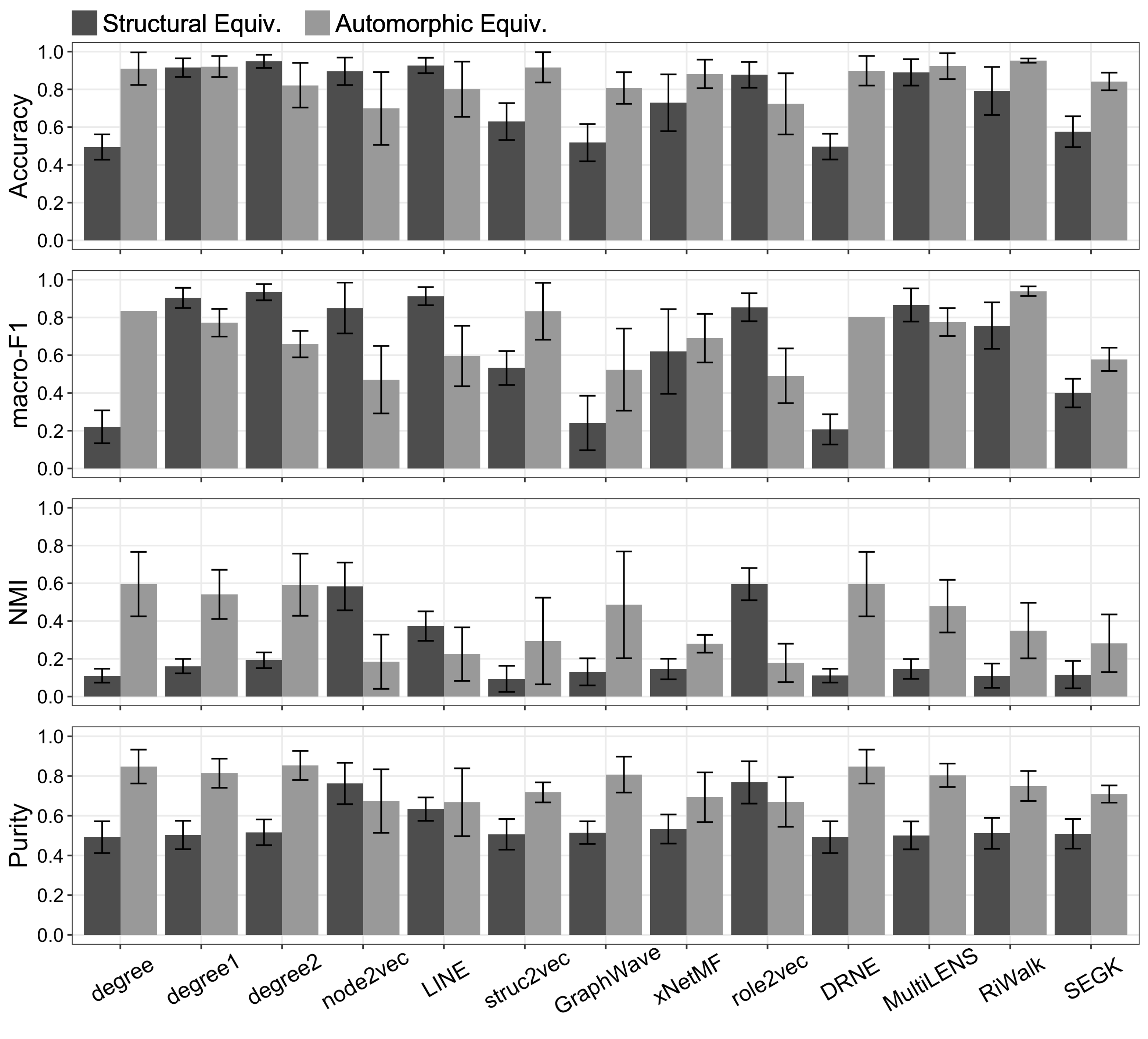}
        \vspace{-0.65cm}
        \caption{Real data}
        \label{fig:downstream-real}
    \end{subfigure}
    \vspace{-0.4cm}
    \caption{Extrinsic evaluation on downstream tasks. Mean and standard deviation is presented for each method on all corresponding synthetic datasets and real datasets for three types of equivalence. Generally, the extrinsic evaluation aligns with the intrinsic evaluation.} 
\label{fig:equivalence-downstream}
\vspace{-0.2cm}
\end{figure*}

\vspace{0.1cm}
In Figure~\ref{fig:equivalence-downstream} we show the results for all three types of equivalence on synthetic (left) and real (right) data.  
For structural and automorphic equivalence evaluation, we use  the enlarged synthetic graphs described in the top section of Table~\ref{tab:enlarged_network_s_a}. Again, we exclude City of Stars for structural equivalence evaluation for the same reason explained in Section~\ref{sec:synthetic_intrinsic_methodology}. For the real data evaluation, we  use the three air-traffic networks and Facebook. For regular equivalence, we use the enlarged synthetic graphs described in the bottom section of Table~\ref{tab:enlarged_network_s_a}.  No real world dataset is appropriate for regular equivalence evaluation as discussed before. 

\subsubsection{Results.}
We generally see similar trends to the intrinsic evaluation.  For example, proximity-based methods node2vec and LINE are generally best at capturing structural equivalence in both real and synthetic datasets, in supervised and unsupervised downstream tasks. They take a backseat to most other methods, however, at predicting automorphic or regular equivalences.  We observe, however, that MultiLENS improves considerably in downstream tasks. 

Differences between methods are often more pronounced in synthetic datasets, which are designed to exhibit highly distinctive structural roles. For instance, LINE and node2vec are over 4$\times$ more accurate at predicting structural equivalence than structural embeddings GraphWave and xNetMF, a gap that remains but shrinks considerably in the real datasets. Similarly, in synthetic datasets, GraphWave and xNetMF achieve near-perfect clustering scores, as do degree distribution features from 1-hop and 2-hop neighborhoods (which perform competitively with other structural embedding methods at capturing equivalences across our extrinsic evaluations).

\begin{observation}
The clear structural roles of our synthetic datasets are a good way to expose differences between structural embedding methods.  
\end{observation}

In general, we observe similar results between intrinsic and extrinsic evaluation as well as synthetic versus real networks.  This suggests that intrinsic evaluation of structural embeddings can \textit{often} be a good proxy of its ability to perform in a downstream task, without adding the additional variable of the downstream machine learning algorithm.  Similarly, synthetic networks that can be manufactured to exhibit distinctive structural roles that are known \emph{a priori} are a good controlled experimental environment for structural node embedding.  However, researchers should be mindful that there may be exceptions to these trends: MultiLENS is one in both cases, performing far better in extrinsic evaluation and on real data.  The word embedding literature has noted that intrinsic evaluations of embeddings may not always accurately predict performance in downstream tasks \cite{chiu2016intrinsic}.  Thus, both forms of analysis are worthwhile to perform.  

\begin{observation}
Intrinsic evaluation and/or synthetic datasets are often a good approximation of a method's performance on graph mining tasks, but are not a complete substitute for extrinsic evaluation on real datasets.
\end{observation}

\vspace{-0.1cm}
\section{Mining a Single Network with Structural Embedding}
\label{sec:real-exp}

\begin{figure*}[t!]
    \centering
    \begin{subfigure}[b]{0.58\textwidth}
        \centering
        \includegraphics[width=.96\linewidth]{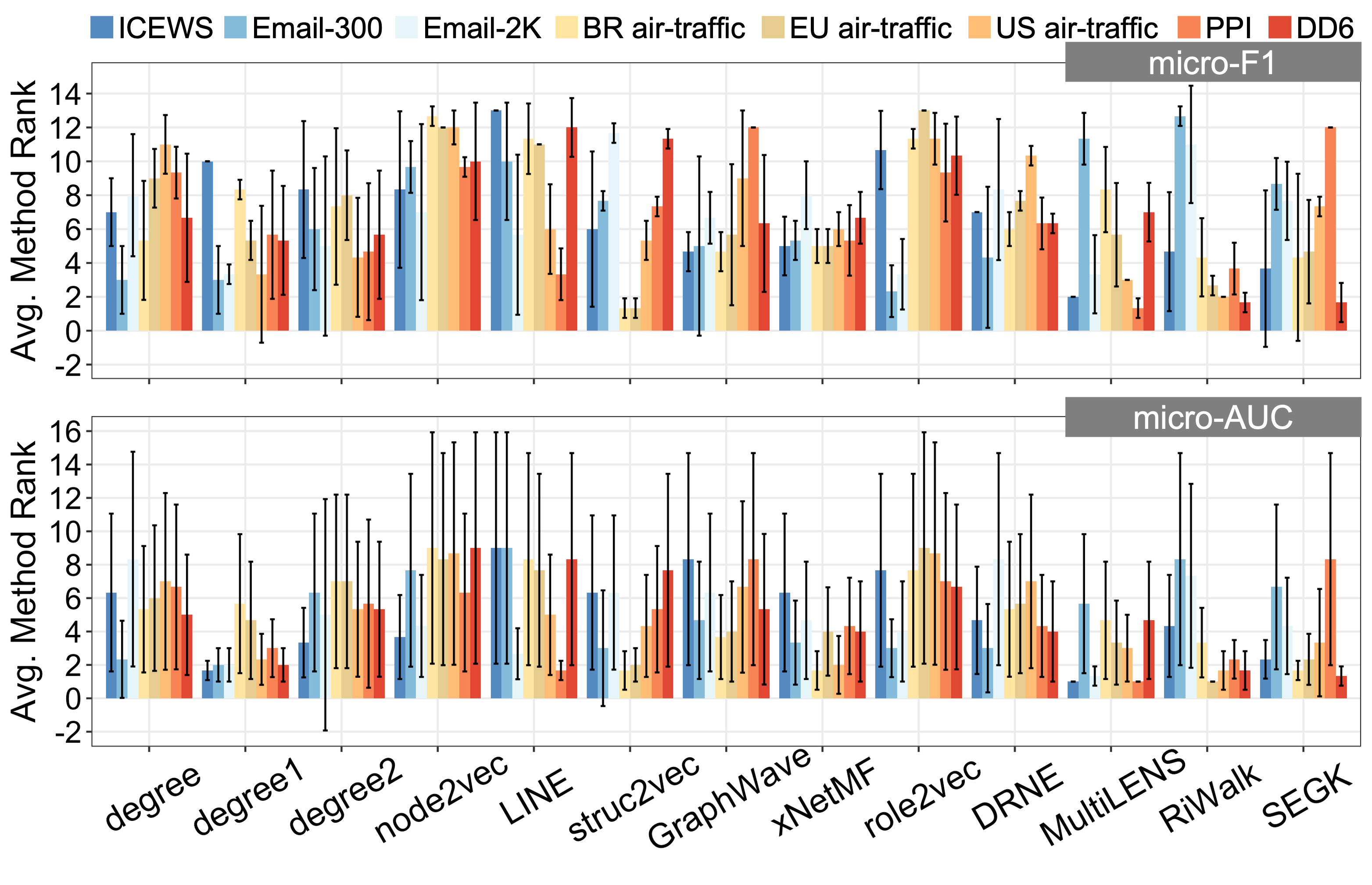}
    \vspace{-0.1cm}
    \caption{Effect of the classifier across the real data: large standard deviations in the embedding rankings over different classifiers show that they may dramatically affect relative performance.}
    \label{fig:effect-classifier}
    \end{subfigure}
    \hfill
    \begin{subfigure}[b]{0.4\textwidth}
        \centering
        \includegraphics[width=.96\linewidth]{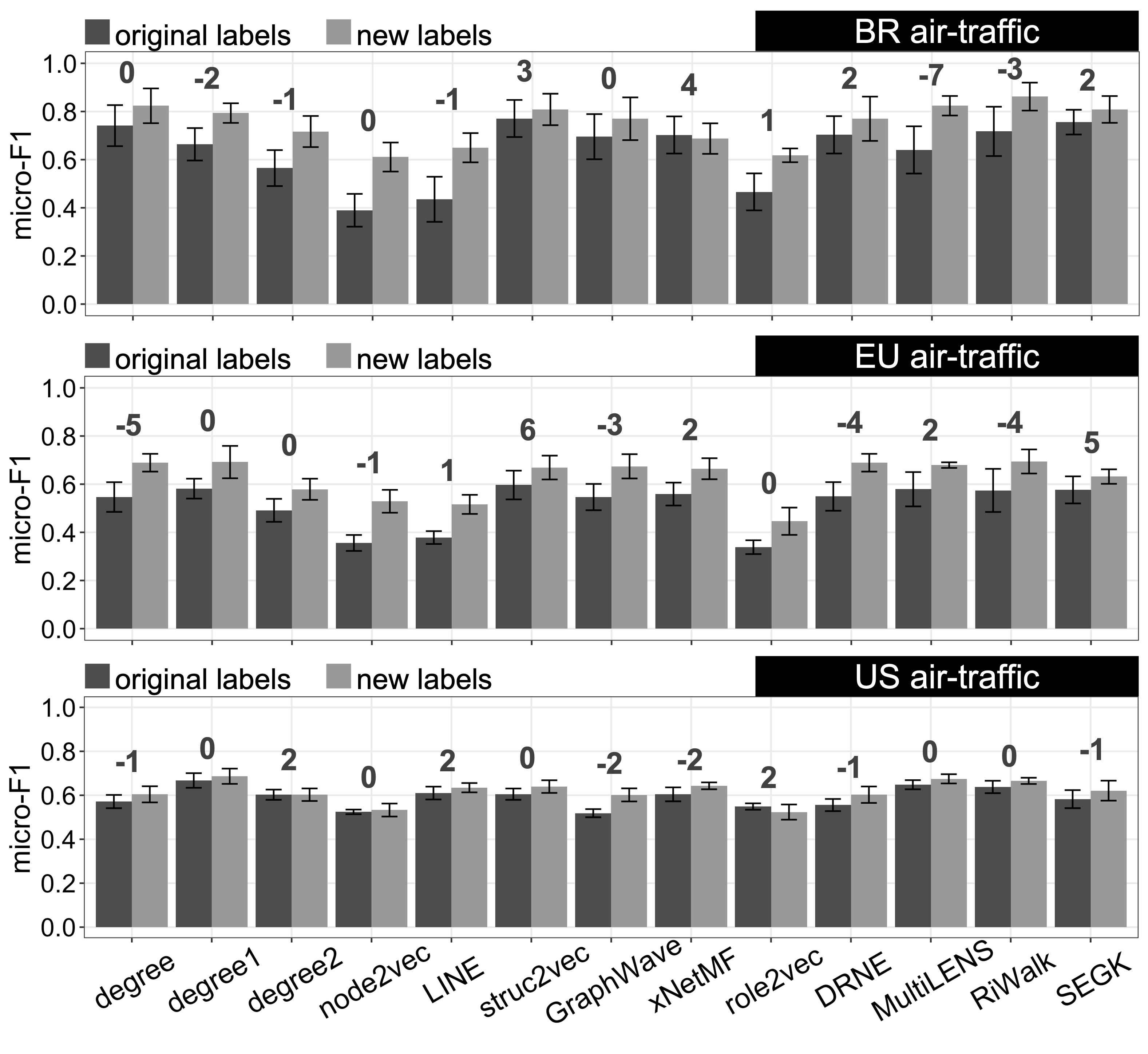}
        \vspace{-0.15cm}
        \caption{Different labeling schemes: Numbers represent decrease in ranking under new labeling.  Most embeddings' rankings change. }
        \label{fig:results-by-label}
    \end{subfigure}
    \vspace{-0.3cm}
    \caption{The performance of different embedding methods in downstream classification tasks heavily depends on the choice of the classifier and the definition of the ground-truth labels.} 
    \vspace{-0.15cm}
\end{figure*}

We now compare methods for structural node embedding on real-world networks and task-specific settings on graph mining tasks with \textit{externally given node labels} (unlike Section~\ref{sec:equiv-labels-extrinsic} that relied on equivalence-defined labels).  We consider single-network tasks, and specifically the task of node classification, 
which can be formulated as a well-studied supervised machine learning problem.  Before presenting comparative results, we identify two important real-world observations that can confound the fair evaluation of structural embeddings on real datasets.  We thus perform analysis of how methods' performance varies as a function of these factors. 

\subsection{Experimental Configuration}
\noindent \textbf{Data}.  We use all the real datasets in Table~\ref{tab:real} except for the BlogCatalog and Facebook datasets, which do not have node labels that reflect structural roles of nodes and as the basis for extrinsic evaluation are usually reserved for proximity-preserving node embeddings.  The remaining datasets all come with node labels, which we use various machine learning classifiers to predict given the features derived from node embedding.   

\vspace{0.1cm}
\noindent \textbf{Classifiers}.  Our classifiers are all popular machine learning models and have been used to evaluate node embeddings on downstream tasks.  Along with their hyperparameter settings, they are:

\begin{itemize}
\item Logistic regression and
linear SVM: these are two commonly used linear models.  We set the parameter $C = 1.0$, and use a one-vs-rest strategy for multiclass classification. The other parameters are set as default from the scikit-learn packages \cite{scikit-learn}.
\item $k$-nearest neighbors ($k$-NN): This classifier arguably provides the purest measurement of the geometry of the embedding space, as no additional learning is provided.   We use $k$ = 5 and Euclidean distance for distance measurement.
\end{itemize}
We introduce any additional protocols specific to a particular experiment as it becomes relevant.  

\subsection{The Effect of the Classifier}
\label{sec:classifier}
Since the structural embedding methods we consider are unsupervised, they are not optimized for performance on a particular downstream task.  Furthermore, we can use any of several different common machine learning metrics to measure task-specific performance.  We now study how these downstream variables affect the assessment of the ``upstream'' embedding methods that are our primary interest.    

\subsubsection{Methodology}
Importantly, we note that the downstream machine learning models used to classify node labels from embeddings can have a significant effect on the results.  We illustrate this point through the use of several classifiers: 
logistic regression, $k$-nearest neighbors, and a linear SVM. In Figure~\ref{fig:effect-classifier}, we report results on all datasets where we average the relative ranks of all of our methods across all classifiers (based on different metrics).  We use two different metrics: (micro)-AUC and F1 score. 

\subsubsection{Results}
We see that there is a considerable standard deviation in the rankings, indicating that with simply using a different downstream machine learning model atop the same embeddings can change the evaluation of which embedding method is ``better.''  We also observe a difference between the two metrics, indicating that different embedding methods may be better or worse depending on the evaluation metric used.  With many different classifiers and metrics being used in the literature, it is important to keep in mind that these too are variables that may affect the performance.

\begin{figure*}[t!]
    \includegraphics[width=\linewidth]{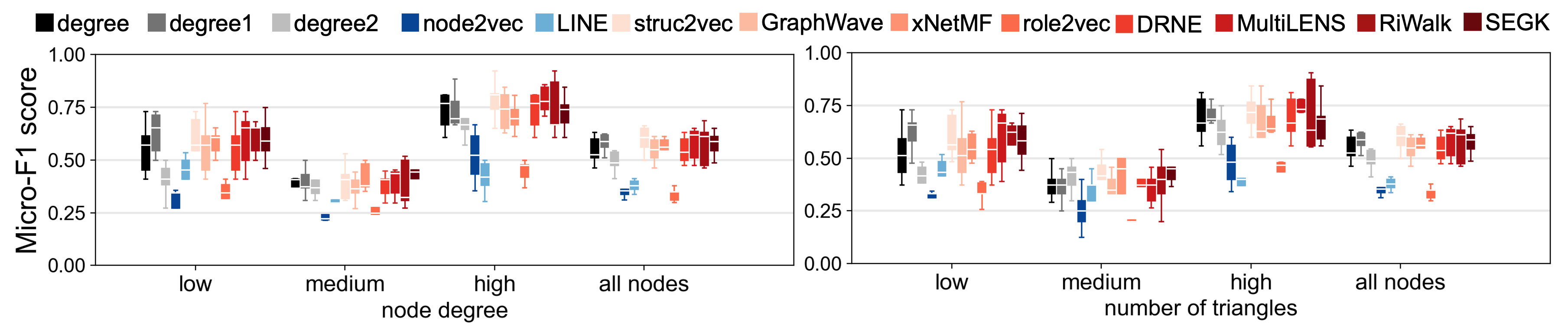}
  \vspace{-0.45cm}
    \caption{[Best viewed in color]
       Performance by node degree and participating triangles on the original label on EU air-traffic: nodes with more ``extreme'' degrees are more accurately classified. 
       Box plot based on 5-fold CV results. 
    } \label{fig:deeper-EU-deg-triangles_original}
  \vspace{-0.35cm}
\end{figure*}

\subsection{The Effect of Label Definitions}
\label{sec:label-dfn}

\subsubsection{Methodology}
In Figure~\ref{fig:results-by-label}, we show the results of different embedding methods on the air-traffic datasets for two different labeling schemes: the original ones resulting in balanced classes, and our relabeling in Section~\ref{sec:limitations}. For brevity, we report Micro-F1 scores obtained using logistic regression, and annotate the decrease in ranking under the new labeling, per method. 

\subsubsection{Results}
We see noticeable differences in performance under the two different labeling methods; In several cases, this can change the comparative ranking of the different methods.  For example, MultiLENS {and RiWalk} are in the middle of the pack under the old labels but the best methods at predicting the new labels.  

Recent works have observed that node classification involves a labeling process that may be uncorrelated with the graph itself, which may complicate evaluation~\cite{epasto2019single}. In these airport datasets, where the labels were arbitrarily discretized, this issue is even more pronounced. 
The fact that two (reasonable) ways of generating node labels can yield different results among structural embedding methods suggest that each structural embedding method best captures certain structural roles in the network, and it then becomes an empirical question how well these roles are correlated with the labels.  (Note that the airport labels are not connected to any particular roles.)  This is the reason why we have performed our previous analysis dissecting the structural role information that each embedding method best captures.   

\begin{observation}
Many factors unconnected to the node embedding process can affect the apparent relative effectiveness of unsupervised structural node embedding methods on downstream graph mining tasks, including:
\begin{itemize}
    \item the downstream machine learning classifier,  
    \item the metric used to evaluate performance, and
    \item the way that node labels are defined.
\end{itemize}
\end{observation}

\subsection{Deeper View Into the Performance Scores}
\label{sec:single-deeperview}

Aggregate performance of a classifier over the whole dataset does not tell the whole story.  It is also worth exploring what kinds of nodes (e.g., high degree) can be most easily classified by the various structural embedding methods.  

\subsubsection{Methodology} 
For degree-based analysis, per dataset with maximum degree $MaxD$, we categorize the nodes into low-degree $[0, MaxD^{\frac{1}{3}})$, medium-degree $[MaxD^{\frac{1}{3}}, MaxD^{\frac{2}{3}})$ and high-degree  $[MaxD^{\frac{2}{3}}, MaxD]$ buckets. We then perform classification 
evaluation per bucket. {We apply the same partitioning methodology for the analysis of participating triangles.}
We use as a case study the EU air-traffic network (we see similar trends in other data). Its maximum degree and maximum number of participating triangles are 202 and 3450, respectively.  
\subsubsection{Results.}
In Figure~\ref{fig:deeper-EU-deg-triangles_original}, 
we observe that in general, all methods perform best at classifying nodes with high connectivity, as measured by either degree and/or participating in a large number of triangles.  This is not surprising and corroborates the literature, as these nodes' local neighborhoods contain richer information~\cite{nandanwar2016structural}.  Slightly more surprisingly, the least-connected nodes are the next easiest to classify.  

\begin{observation}
Current structural embedding methods are most effective at distinguishing ``extreme'' network positions in the latent feature space compared to moderate ones.  
\end{observation}

  Some network positions are easy to identify.  For instance, simply using the node degree as a feature (\texttt{degree}) performs best at classifying high degree nodes, but is less effective at classifying low- and medium-degree nodes even compared to \texttt{degree1}, where neighbors' degrees are considered as features.  In general, however, relative ranks of methods are fairly well-preserved across buckets.   

\subsection{A Comprehensive Embedding Comparison: Single-Network Tasks}

\begin{figure}[t!]
    \centering
    \includegraphics[width=.9\textwidth]{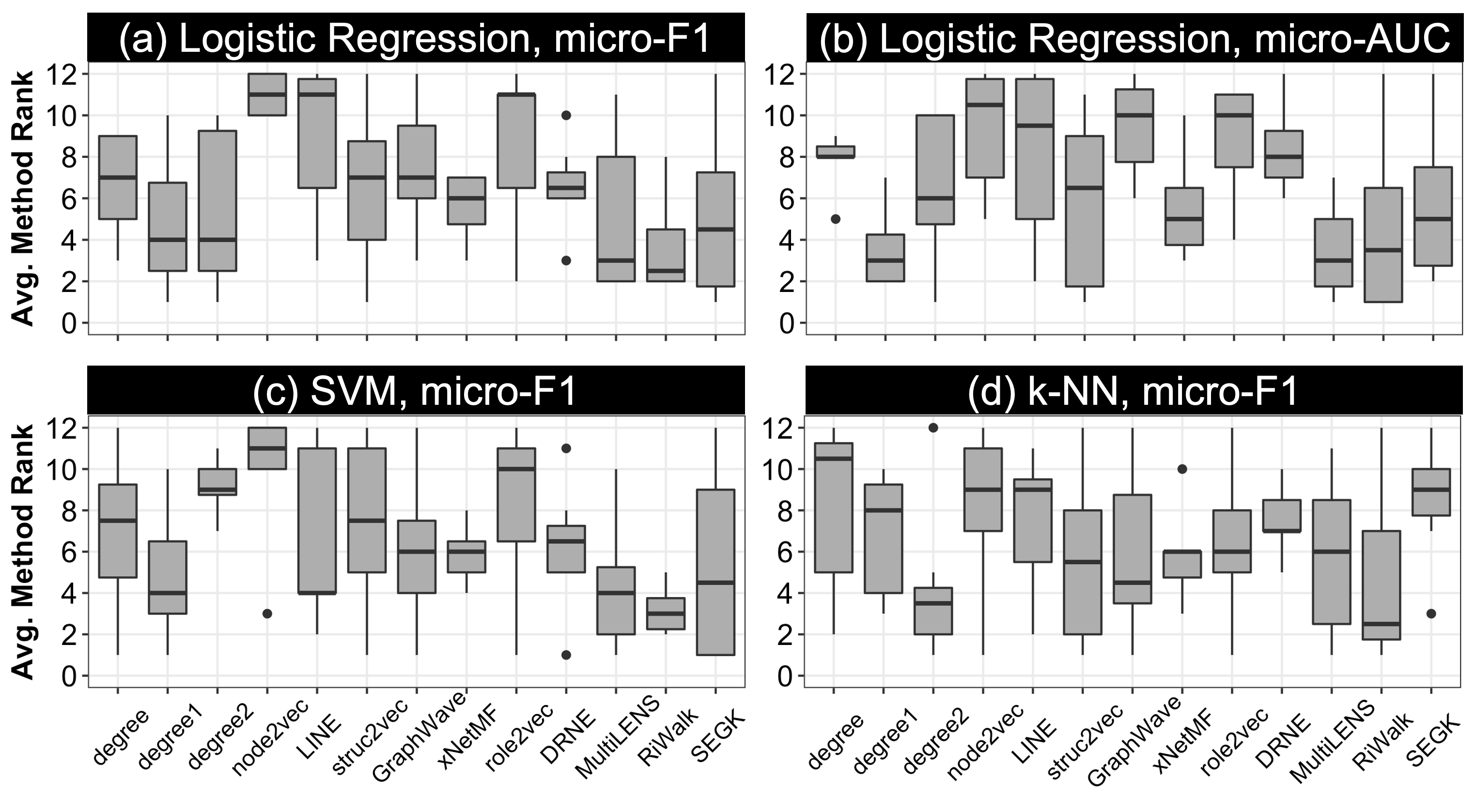}
    \caption{Lower is better: performance summarized across all the real datasets.  While there is no clear winner, methods based on local degree distribution tend to be consistently top performers.}
    \label{fig:results-by-classifier}
\end{figure}

Having carefully considered the effects of several external factors, we now offer a more comprehensive comparison of embedding methods in Figure~\ref{fig:results-by-classifier}: we give their general rankings (lower is better) per classifier and metric across all real datasets. 
We observe that there is no clear winner of an embedding method, particularly as datasets, labels, classifiers, and metrics may all change.  However, we can see that node embedding methods designed to preserve proximity in the network---node2vec and LINE---generally have poorer rankings, as is to be expected for a task where the nodes' structural role carries most of the signal.  Other methods are more mixed: e.g., GraphWave achieves a more competitive ranking by both metrics displayed with an SVM and less so with logistic regression.  

Some of the best-ranking methods across the board are MultiLENS, SEGK and variations of our degree distribution features.  Significantly, they all share common design choices, explicitly modeling a node's position within a local neighborhood using degree-based connectivity (after one iteration, the Weisfeiler-Lehman graph kernel used by SEGK gives nodes the same label if and only if they have same degree, in the absence of other node label information). We believe that the expressive power of local degree distributions has strong implications for future work in structural embedding, as a baseline and an inspiration for methodological design.  

\begin{observation}
Current individual network mining tasks depending on the structural roles of nodes can be solved effectively with local aggregation of degree-based connectivity information.
\end{observation}

\vspace{-0.1cm}
\section{Mining Multiple Networks with Structural Embedding}
\label{sec:multinetwork}
One important benefit of structural embedding methods is that they can be used to compare nodes across graphs~\cite{rolx, xnetmf, rgm}.  In this section, we apply different structural embeddings to two graph mining tasks involving cross-network comparison: network alignment, which finds node-level matchings between different graphs, and graph classification, which involves comparing entire graphs.  

\subsection{Network Alignment}
\subsubsection{Methodology}
Network alignment can be formulated as nearest-neighbor search given comparable structural node embeddings~\cite{xnetmf}. We follow established procedures for simulating a network alignment problem with known ground truth correspondences~\cite{xnetmf}: we align a graph with adjacency matrix $\matA$ to a randomly permuted version of itself given by $\matP \adj \matP^\top$ for random permutation matrix $\matP$, to which we add noise by removing edges with probability $p$.  We use a $k$-d tree to quickly match all the nodes in one graph to the nearest neighbor in another graph by embedding similarity, and compute the resulting accuracy. We perform this experiment on the two datasets used in previous works~\cite{xnetmf} and described in Section~\ref{sec:data}, using 1\% and 5\% noise, the lowest and highest noise levels considered in~\cite{xnetmf}.

\subsubsection{Results}
\begin{wrapfigure}{R}{0.5\columnwidth}
    \centering
    \includegraphics[width = 1.0\linewidth]{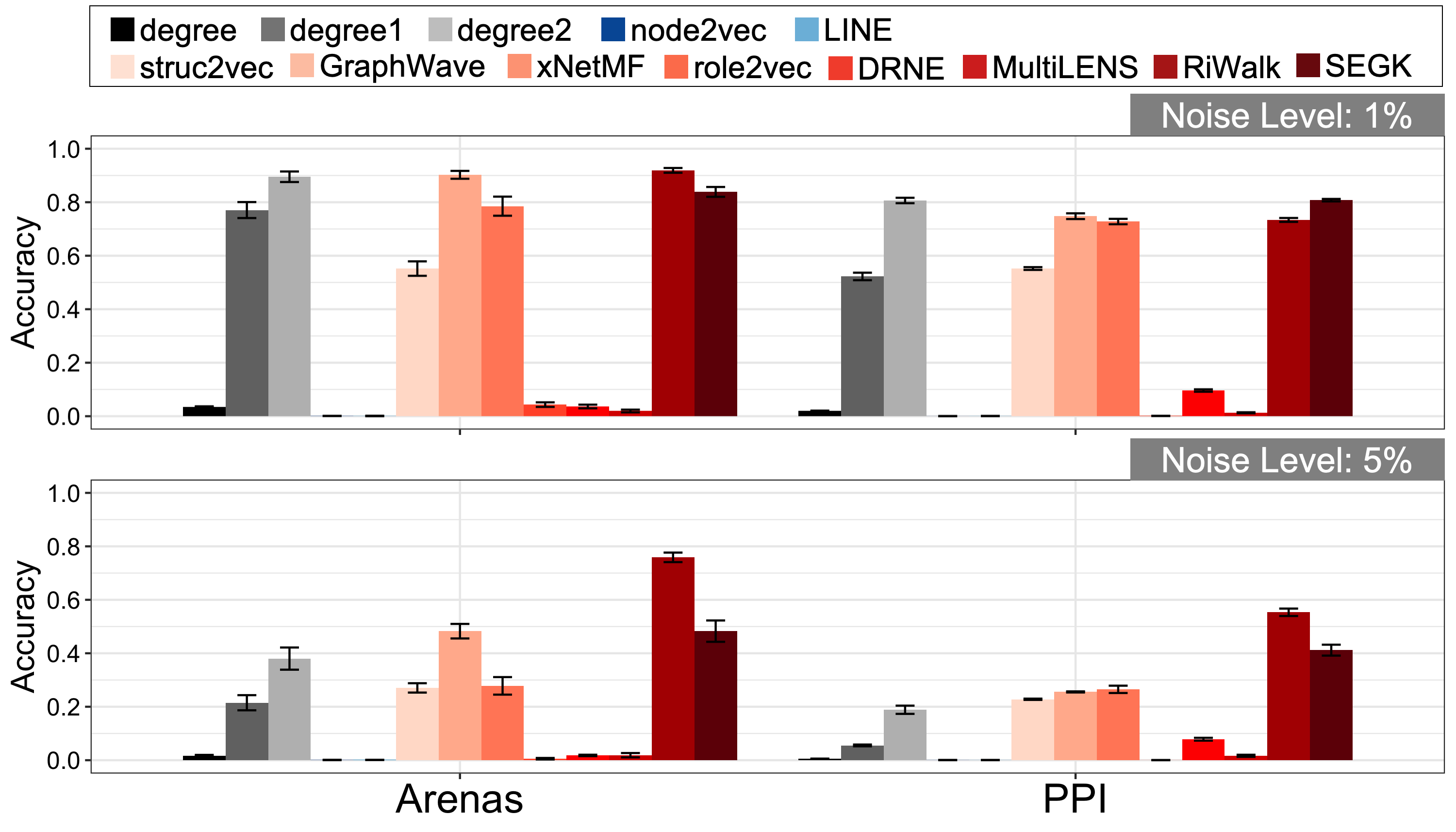}
    \caption{Graph alignment results.}
    \label{fig:multinetwork-alignment}
\end{wrapfigure}

In Figure~\ref{fig:multinetwork-alignment}, we see that xNetMF, which was originally proposed for graph alignment, captures cross-network node similarities.  Proximity-preserving node embedding methods LINE and node2vec are unable to succeed on this task.  Neither are role2vec, DRNE or MultiLENS, which may be regarded as hybrids of proximity-preserving and structural embeddings.  Node degree alone is too weak a structural descriptor to meaningfully align nodes (many nodes in a network have the same degree), but degree distributions of higher-order local neighborhoods (2nd-order is always better than first-order) are also sufficiently expressive structural descriptors to perform on par with xNetMF (which also preserves distributional information of neighbor degrees) in many cases.  

For this task, some of the most successul methods are successors of xNetMF: SEGK and RiWalk.  Both methods generalize the structural connectivity measure between nodes beyond degree alone, which RiWalk notes can be ambiguous~\cite{riwalk}.  In particular, both methods use the Weisfeiler-Lehman neighborhood aggregation method, a well-known heuristic for graph-level similarity which has its roots in a graph isomorphism test~\cite{wlgk}.  The neighborhood aggregation process iteratively relabels each node, capturing degree-based statistics in early iterations but gradually building up higher-order information.  

Especially in the more challenging alignment settings with 5\% noise, RiWalk performs better than all other methods, even SEGK.  One reason for this may be because RiWalk is not restricted to local neighborhoods, while methods like SEGK (and xNetMF, struc2vec) model only $k$-hop neighborhoods of each nodes. Concurrent method GraphWave is also successul (close to SEGK and ahead of xNetMF on the Arenas dataset); GraphWave also considers patterns of local connectivity using heat diffusion processes rather than degree.  We note, however, that a partial explanation of the success of structural embedding methods that do not explicitly model node degree may be in part due to the noise model~\cite{xnetmf}.  The removal of edges may affect the degree distribution more obviously than it affects diffusion processes on graphs.  

\begin{figure*}[t!]
    \centering
    \includegraphics[width=\textwidth]{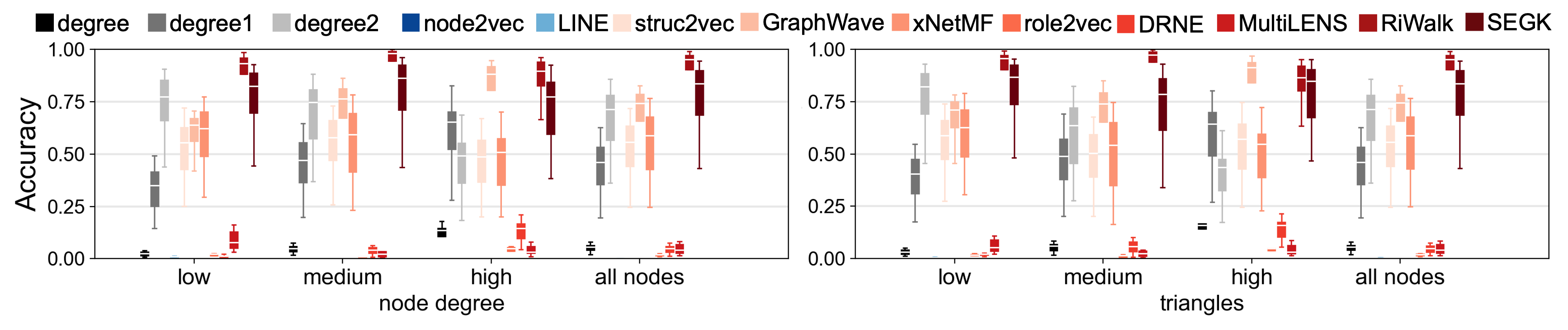}
    \vspace{-0.45cm}
    \caption{[Best viewed in color] Deeper view into performance scores for network alignment.}
    \label{fig:alignment-deeperview}
    \vspace{-0.35cm}
\end{figure*}

In Figure~\ref{fig:alignment-deeperview}, we take a deeper look into the alignment results at 1\% noise on a node-level basis, as in Section~\ref{sec:single-deeperview}.  Specifically, after grouping nodes by degree or number of participating triangles, we plot how accurately each method aligns nodes in each group.  We see that a few methods like \texttt{degree1} and GraphWave are best at classifying high-connectivity nodes, which as noted in Section~\ref{sec:single-deeperview} have richer information in their local neighborhoods.  On the other hand, many methods see a decline for high-connectivity models, which may be due in part to the noise generation model; high-degree nodes in the original graph may have no counterpart with a similar degree in the corrupted graph, where edges have been removed and the node degrees are reduced in expectation.  However, the best-performing methods, such as RiWalk, are fairly consistent across levels of connectivity; their lack of explicit dependence on node degree allows them to be more robust to the noise model, but they are still expressive enough to capture sufficient local structural information even for poorly connected nodes.  

\begin{observation}
For alignment of noisy networks, degree-based connectivity alone can be brittle, and the most robust methods generalize the notion of node connectivity beyond degree.  
\end{observation}

\subsection{Network Classification}
\subsubsection{Methodology}
To classify networks from the structural embeddings of nodes, we use RGM, an unsupervised graph feature map that captures the distribution of the node embeddings in feature space~\cite{rgm}.  We then train a linear SVM on the resulting graph features.   RGM was shown to work with different choices of node embeddings and yielded comparable or better accuracy to a large variety of baseline graph kernels, neural networks, and unsupervised feature construction methods at faster runtime~\cite{rgm}.  We use recommended settings of 4 levels of resolution and 2 iterations of Weisfeiler-Lehman label expansion (when no node labels are available, we begin this process with uniform labels~\cite{rgm}) for RGM.  This label expansion helps RGM aggregate the node embeddings more accurately, but we do not use node labels during embedding.  For the downstream classification, we consider a linear SVM, as it was shown that RGM with a linear SVM approximates a kernel machine~\cite{rgm}.

\subsubsection{Results}
\begin{wrapfigure}{R}{0.73\columnwidth}
    \centering
    \includegraphics[width = 1.0\linewidth]{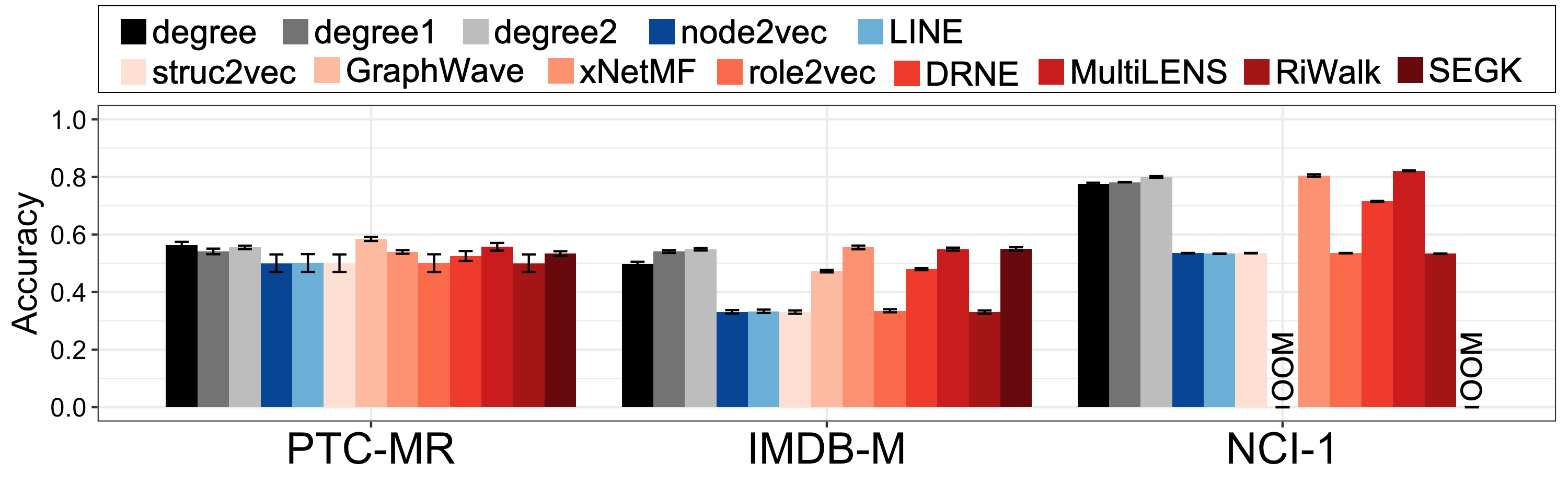}
    \caption{Graph classification results.  Embedding methods modeling local neighborhoods tend to do best.}
    \label{fig:multinetwork-classification}
\end{wrapfigure}

We plot the results for the different embedding methods in Figure~\ref{fig:multinetwork-classification}.   For ease of inspection, we also report the numbers in tabular format (Table~\ref{tab:graphclassification-results}).  We also give additional context relative to competing methods representing other families of techniques, by including results from the state-of-the-art Weisfeiler-Lehman subtree graph kernel~\cite{wlgk} along with GIN~\cite{gin}, a state-of-the-art graph neural network (we use the numbers for the best-performing GIN-0 variant reported in the original paper~\cite{gin}).  

\begin{table*}[t]
\caption{Accuracy of various structural embeddings used in the RGM framework~\cite{rgm} for graph classification, plus strong baselines from other graph classification  techniques.  
(OOM = Out of Memory.) Most accurate embedding method in RGM marked in bold.  Average rank computed by accuracy, with ties broken by standard deviation if applicable.  Tied methods given rank of highest tie, OOM given a rank below all methods that completed.  
}
\label{tab:graphclassification-results}
\centering
\vspace{-0.3cm}
{\small
    \begin{tabular}{l@{\hskip 0.3in}ccc|c}

     \\ \toprule 
     
    {\bf Method} &
    \multicolumn{1}{c}{\bf PTC-MR} & 
    \multicolumn{1}{c}{\bf IMDB-M} & 
    \multicolumn{1}{c}{\bf NCI1} &
    \multicolumn{1}{|c}{\bf Average Rank} 
    \\ \midrule

    \mytop {\bf degree} 
    & 56.3 $\pm$ 1.1 
    & 49.7 $\pm$ 0.9 
    & 77.5 $\pm$ 0.4 
    & 4.33 \\
    
    \mytop {\bf degree1}  
    & 54.1 $\pm$ 1.0 
    & 54.0 $\pm$ 0.5 
    & 78.2 $\pm$ 0.1 
    & 4.67 \\

    \mytop {\bf degree2}  
    & 55.5 $\pm$ 0.6 
    & 54.9 $\pm$ 0.4 
    & 80.0 $\pm$ 0.3 
    & 3.33 \\ 

     \mytop {\bf node2vec}  
    & 50.0 $\pm$ 3.0 
    & 33.1 $\pm$ 0.6 
    & 53.5 $\pm$ 0.1 
    & 9.67 \\ 
    
     \mytop {\bf LINE}  
    & 50.1 $\pm$ 3.1 
    & 33.3 $\pm$ 0.6 
    & 53.5 $\pm$ 0.1 
    & 8.67 \\ 
    
    \mytop {\bf struc2vec}  
    & 50.0 $\pm$ 3.0 
    & 33.0 $\pm$ 0.6 
    & 53.5 $\pm$ 0.1 
    & 10 \\ 
    
    \mytop {\bf GraphWave}  
    & \textbf{58.5 $\pm$ 0.7} 
    & 47.2 $\pm$ 0.4 
    & OOM 
    & 7 \\
    
     \mytop {\bf xNetMF}  
    & 53.9 $\pm$ 0.6 
    & \textbf{55.5 $\pm$ 0.7} 
    & 80.5 $\pm$ 0.4 
    & 3 \\ 
    
     \mytop {\bf role2vec}  
    & 50.1 $\pm$ 3.1 
    & 33.5 $\pm$ 0.5 
    & 53.5 $\pm$ 0.1 
    & 8.33 \\ 
    
     \mytop {\bf DRNE}  
    & 52.6 $\pm$ 1.7 
    & 47.9 $\pm$ 0.4 
    & 71.5 $\pm$ 0.2 
    & 7 \\ 
    
     \mytop {\bf MultiLENS}  
    & 55.7 $\pm$ 1.3 
    & 54.9 $\pm$ 0.5 
    & \textbf{82.1 $\pm$ 0.1} 
    & \textbf{2.67} \\ 
    
     \mytop {\bf RiWalk}  
    & 50.0 $\pm$ 3.0 
    & 33.0 $\pm$ 0.6 
    & 53.5 $\pm$ 0.1 
    & 10 \\ 
    
      \mytop {\bf SEGK}  
    & 53.3 $\pm$ 0.8 
    & 55.0 $\pm$ 0.6 
    & OOM 
    & 7 \\
    
    \bottomrule 
    \end{tabular}
}
\vspace{-0.2cm}
\end{table*}

We see that particularly on the social networks dataset IMDB-M, skip-gram based methods whose context sampling is not locally restricted (node2vec, role2vec, RiWalk) yield poor performance.  node2vec's performance is also explained by the fact that node proximity information does not lead to comparable representations between different graphs, as is confirmed by~\cite{rgm} and by the poor performance of LINE.  However, RiWalk and role2vec's struggles may be because on these graphs, the random walks oversample the graph and wash out distinguishing structural information.  

\begin{observation}
For network classification, sampling structural context with random walks risks blurring too much structural information on the small graphs commonly used as benchmarks.  
\end{observation}

This generalizes the finding in~\cite{rgm} that methods such as node2vec and struc2vec perform poorly.  There, the explanation was that such methods were not inductive; we see that this is true, as LINE, which does not use random walks but does depend on a transductive notion of proximity, also performs equally poorly.  However, even structural embedding methods like RiWalk and role2vec, which we saw were useful for cross-network tasks like network alignment, perform poorly here: indicating that the mechanism they all use to sample context may be at fault.  Note that the more memory-intensive baselines GraphWave and SEGK are unable to run on the largest NCI1 dataset.  

On the other hand, the best performing methods are those that explicitly model local neighborhoods: xNetMF and SEGK. 
This corroborates the finding that local structural information is sufficient on many existing graph classifiation benchmarks~\cite{cai2019simple}. Degree variants also do well, with higher-order hop distances achieving more accuracy on IMDB-M and NCI1.  However, on the PTC-MR dataset, which consists of smaller molecular graphs that may not contain the complex structural roles arising from social behavior, we see less of a gap between all methods, and in fact of the three degree-based methods, \texttt{degree} does best (by a marginal amount).  This indicates that on this dataset, extremely limited local structural information is sufficient.  

\begin{observation}
For network classification, the best methods locally model the connectivity of each node.  Considering each node's higher-order connectivity does slightly improve performance on medium to large datasets.  
\end{observation}

As an aside, while it is not our goal to set a task-specific state of the art, within RGM the structural embeddings yield competitive performance to other leading methods.  Compared to results from the state of the art graph isomorphism networks reported in ~\cite{gin} and Weisfeiler-Lehman graph kernels~\cite{wlgk} reported in ~\cite{rgm}, the best embedding-based methods yield clearly higher numbers on IMDB-M and trail by a fraction of a percentage point on NCI1 (they trail further on PTC-MR).  Note that our feature learning method is completely unsupervised (unlike the GIN neural network) and we do not tune the parameters (e.g. number of binning levels) of RGM, which could further improve performance.  

\subsection{A Comprehensive Embedding Comparison: Multi-Network Tasks}
For graph classification, the results resemble the results from the single-network tasks in Section~\ref{sec:real-exp}.  The best methods aggregate local connectivity information for each node; these include xNetMF, MultiLENS, SEGK (when it is able to run), and variants of the local degree histograms. On the large datasets, considering second-order neighbors for each node improves over considering only the nodes' features or that of its immediate neighbors, indicating that modeling higher-order connectivity does somewhat help for this task.

For graph alignment, generalizing node connectivity beyond degree is helpful, which is why the most successul methods are RiWalk, SEGK, and GraphWave.  This may be in part because the noise model of edge removal~\cite{xnetmf} throws off the degree distribution of the graphs, making the degrees in the noisy graph slightly lower.  RiWalk and GraphWave are not explicitly confined to modeling any $k$-hop neighborhood, but SEGK is.  This implies that modeling local structural information does not necessarily hurt performance, but using a statistic like degree to assess structural identity that is particularly brittle under the common noise model is more likely to lower a structural embedding method's performance on network alignment, particularly when the noise is higher.

\begin{observation}
Multi-network tasks can be solved using node embeddings that capture local structural information.  For graph classification, degree is a sufficiently expressive measure of connectivity, but graph alignment requires a more generalized measurement of connectivity.
\end{observation}

\section{Discussion and Conclusions}
\label{sec:conclusions}

\vspace{-0.1cm}
We conducted a comprehensive empirical study 
to gain a better understanding of the \textit{equivalence} of the nodes in the networks within the context of embeddings. 
Our  study of the various sociological equivalences confirms that structural equivalence is best captured by proximity-preserving embedding methods like node2vec and LINE, as its definition implies despite its name.  On the other hand, methods like struc2vec, xNetMF and GraphWave perform well in automorphic and regular equivalence (though the definition of the latter depends on edge types and is challenging to define in a principled way without this information). 

We have split our analysis into two parts (Section~\ref{sec:emb-equivalences}): intrinsic evaluation, which explores the relationship of nodes' embedding similarities and other measures of similarity given by sociological equivalence, and extrinsic evaluation of the embeddings' performance in the context of downstream tasks such as classification or clustering. Our work is one of the first to perform intrinsic \textit{and} extrinsic evaluation of node embeddings (either structural or proximity-based).  

While we largely observe similar performance trends in intrinsic and extrinsic evaluation, we also notice some inconsistent trends, a phenomenon which has also been observed in word embedding~\cite{chiu2016intrinsic}.  For example, MultiLENS is far from a standout in intrinsic evaluation but a top runner in extrinsic evaluation.  
In both intrinsic and extrinsic clustering evaluation, we have found a complex relationship between the distance metric used (cosine or Euclidean) and the results, which perhaps surprisingly is not always consistent with the metric used in the various embedding objectives.  

More generally, we have found that the performance of structural node embedding methods is highly sensitive to many factors that are often chosen seemingly arbitrarily: choice of classifier, performance metric, or node labeling method (Section~\ref{sec:real-exp}).  Changing any of these can not only change methods' absolute performance but also their rankings relative to each other, arbitrarily making one embedding method appear better or worse than another.  
Comparing comprehensively across classifiers, performance metrics, datasets, and labeling schemes, we see that the simple structural property, node degree, can be the building block for some of the most effective methods.  Our \emph{local degree histograms} are a simple baseline that proves surprisingly effective across all of our experiments.  They may inspire the design of future methods: indeed, they are highly related to xNetMF and MultiLENS, two existing embedding methods that also exhibit generally strong performance.   

Overall, we hope that our findings can influence the design of further node embedding methods and also pave the way for future evaluation of existing methods. 
With new node embedding methods being developed at a breakneck pace, proper evaluation will, as the word embedding community has found, be essential to progress.

\section*{Acknowledgements}
This material is based upon work supported by 
the NSF under Grant No. IIS 1845491, Army Young Investigator Award No. W911NF1810397, 
an Adobe Digital Experience research faculty award, and Amazon, Facebook and Google faculty awards.
Any opinions, findings, conclusions or recommendations 
expressed in this material are those of the {authors} and do not necessarily reflect the views of the 
funding parties.

\enlargethispage{\baselineskip}
\balance
\bibliographystyle{plain}
\bibliography{BIB/abbrev,BIB/bibliography,BIB/references,BIB/all}

\begin{thebibliography}{10}

\bibitem{DRNE_repo}
{DRNE codebase}.
\newblock \url{github.com/tadpole/DRNE}.

\bibitem{graphwave_repo}
{GraphWave codebase}.
\newblock \url{github.com/snap-stanford/graphwave}.

\bibitem{LINE_repo}
{LINE codebase}.
\newblock \url{github.com/tangjianpku/LINE}.

\bibitem{multilens_repo}
{MultiLENS codebase}.
\newblock \url{github.com/GemsLab/MultiLENS}.

\bibitem{node2vec_repo}
node2vec codebase.
\newblock \url{github.com/aditya-grover/node2vec}.

\bibitem{riwalk_repo}
{RiWalk codebase}.
\newblock \url{github.com/maxuewei2/RiWalk}.

\bibitem{role2vec_repo}
role2vec codebase.
\newblock \url{github.com/benedekrozemberczki/role2vec}.

\bibitem{segk_repo}
{SEGK codebase}.
\newblock \url{github.com/giannisnik/segk}.

\bibitem{struc2vec_repo}
struc2vec codebase.
\newblock \url{github.com/leoribeiro/struc2vec}.

\bibitem{VS_repo}
{Vertex Similarity codebase}.
\newblock \url{github.com/tadpole/DRNE/blob/master/test/test_VS.py}.

\bibitem{xnetmf_repo}
{xNetMF codebase}.
\newblock \url{github.com/GemsLab/REGAL}.

\bibitem{ucinet}
S.P. Borgatti, M.~G. Everett, and L.~C. Freeman.
\newblock {UCINET} 6 for {W}indows: {Software for Social Network Analysis.
  Harvard, MA, Analytic Technologies}, 2002.

\end{thebibliography}


\providecommand{\noopsort}[1]{}
\begin{thebibliography}{10}

\bibitem{role2vec}
Nesreen~K. Ahmed, Ryan~A. Rossi, John~Boaz Lee, Theodore~L. Willke, Rong Zhou,
  Xiangnan Kong, and Hoda Eldardiry.
\newblock role2vec: Role-based network embeddings.
\newblock In {\em DLG KDD}, 2019.

\bibitem{bakarov2018survey}
Amir Bakarov.
\newblock A survey of word embeddings evaluation methods.
\newblock {\em arXiv preprint arXiv:1801.09536}, 2018.

\bibitem{BelthBK20}
Caleb Belth, Alican B\"{u}y\"{u}k\c{c}ak{\i}r, and Danai Koutra.
\newblock A hidden challenge of link prediction: Which pairs to check?
\newblock In {\em IEEE International Conference on Data Mining (ICDM)}, 2020.

\bibitem{BorgattiE92_equiv}
Stephen Borgatti and Martin Everett.
\newblock Notions of position in social network analysis.
\newblock {\em Sociological Methodology}, 22, 01 1992.

\bibitem{CATREGE}
Stephen~P. Borgatti and Martin~G. Everett.
\newblock Two algorithms for computing regular equivalence.
\newblock {\em Social Networks}, 15(4):361 -- 376, 1993.

\bibitem{BorgwardtK05}
Karsten~M. Borgwardt and Hans-Peter Kriegel.
\newblock Shortest-path kernels on graphs.
\newblock In {\em ICDM}. IEEE, 2005.

\bibitem{icews}
Elizabeth Boschee, Jennifer Lautenschlager, Sean O'Brien, Steve Shellman, and
  James Starz.
\newblock {ICEWS Automated Daily Event Data}, 2018.

\bibitem{CONCOR}
{Ronald L} Breiger, {Scott A.} Boorman, and Phipps Arabie.
\newblock An algorithm for clustering relational data with applications to
  social network analysis and comparison with multidimensional scaling.
\newblock {\em J. Math. Psychol.}, 12(3):328--383, 1975.

\bibitem{ppi}
Bobby-Joe Breitkreutz, Chris Stark, Teresa Reguly, Lorrie Boucher, Ashton
  Breitkreutz, Michael Livstone, Rose Oughtred, Daniel~H Lackner, J{\"u}rg
  B{\"a}hler, Valerie Wood, et~al.
\newblock The biogrid interaction database: 2008 update.
\newblock {\em Nucleic acids research}, 36(suppl 1):D637--D640, 2008.

\bibitem{cai2019simple}
Chen Cai and Yusu Wang.
\newblock A simple yet effective baseline for non-attributed graph
  classification.
\newblock In {\em ICLR RLGM Workshop}, 2019.

\bibitem{chiu2016intrinsic}
Billy Chiu, Anna Korhonen, and Sampo Pyysalo.
\newblock Intrinsic evaluation of word vectors fails to predict extrinsic
  performance.
\newblock In {\em Proceedings of the 1st workshop on evaluating vector-space
  representations for NLP}, pages 1--6, 2016.

\bibitem{dalmia2018towards}
Ayushi Dalmia and Manish Gupta.
\newblock Towards interpretation of node embeddings.
\newblock In {\em Companion Proceedings of the The Web Conference 2018}, pages
  945--952, 2018.

\bibitem{graphwave}
Claire Donnat, Marinka Zitnik, David Hallac, and Jure Leskovec.
\newblock Learning structural node embeddings via diffusion wavelets.
\newblock In {\em KDD}, volume~24, 2018.

\bibitem{epasto2019single}
Alessandro Epasto and Bryan Perozzi.
\newblock Is a single embedding enough? learning node representations that
  capture multiple social contexts.
\newblock In {\em WebConf}, pages 394--404, 2019.

\bibitem{EVERETT198877}
Martin~G. Everett and Steve Borgatti.
\newblock Calculating role similarities: An algorithm that helps determine the
  orbits of a graph.
\newblock {\em Social Networks}, 10(1):77 -- 91, 1988.

\bibitem{gemsurvey}
Palash Goyal and Emilio Ferrara.
\newblock Graph embedding techniques, applications, and performance: A survey.
\newblock {\em Knowledge-Based Systems}, 151:78--94, 2018.

\bibitem{goyal2019benchmarks}
Palash Goyal, Di~Huang, Ankita Goswami, Sujit~Rokka Chhetri, Arquimedes Canedo,
  and Emilio Ferrara.
\newblock Benchmarks for graph embedding evaluation.
\newblock {\em arXiv preprint arXiv:1908.06543}, 2019.

\bibitem{node2vec}
Aditya Grover and Jure Leskovec.
\newblock node2vec: Scalable feature learning for networks.
\newblock In {\em KDD}. ACM, 2016.

\bibitem{gurukar2019network}
Saket Gurukar, Priyesh Vijayan, Aakash Srinivasan, Goonmeet Bajaj, Chen Cai,
  Moniba Keymanesh, Saravana Kumar, Pranav Maneriker, Anasua Mitra, Vedang
  Patel, et~al.
\newblock Network representation learning: Consolidation and renewed bearing.
\newblock {\em arXiv preprint arXiv:1905.00987}, 2019.

\bibitem{graphsage}
Will Hamilton, Zhitao Ying, and Jure Leskovec.
\newblock Inductive representation learning on large graphs.
\newblock In {\em NeurIPS}, 2017.

\bibitem{heimann2020structural}
Mark Heimann, Goran Murić, and Emilio Ferrara.
\newblock Structural node embedding in signed social networks: Finding online
  misbehavior at multiple scales.
\newblock In {\em Complex Networks}, 2020.

\bibitem{rgm}
Mark Heimann, Tara Safavi, and Danai Koutra.
\newblock Distribution of node embeddings as multiresolution features for
  graphs.
\newblock In {\em 2019 IEEE International Conference on Data Mining (ICDM)}.
  IEEE, 2019.

\bibitem{xnetmf}
Mark Heimann, Haoming Shen, Tara Safavi, and Danai Koutra.
\newblock Regal: Representation learning-based graph alignment.
\newblock In {\em CIKM}. ACM, 2018.

\bibitem{rolx}
Keith Henderson, Brian Gallagher, Tina Eliassi-Rad, Hanghang Tong, Sugato Basu,
  Leman Akoglu, Danai Koutra, Christos Faloutsos, and Lei Li.
\newblock Rolx: structural role extraction \& mining in large graphs.
\newblock In {\em KDD}, 2012.

\bibitem{node2bits}
Di~Jin, Mark Heimann, Ryan~A. Rossi, and Danai Koutra.
\newblock Node2bits: Compact time- and attribute-aware node representations for
  user stitching.
\newblock In {\em PKDD}, 2019.

\bibitem{ember}
Di~Jin, Mark Heimann, Tara Safavi, Mengdi Wang, Wei Lee, Lindsay Snider, and
  Danai Koutra.
\newblock Smart roles: Inferring professional roles in email networks.
\newblock In {\em KDD}, 2019.

\bibitem{multilens}
Di~Jin, Ryan~A Rossi, Eunyee Koh, Sungchul Kim, Anup Rao, and Danai Koutra.
\newblock Latent network summarization: Bridging network embedding and
  summarization.
\newblock In {\em KDD}, 2019.

\bibitem{kipf2017semi}
Thomas~N. Kipf and Max Welling.
\newblock Semi-supervised classification with graph convolutional networks.
\newblock In {\em ICLR}, 2017.

\bibitem{koblenz}
J{\'e}r{\^o}me Kunegis.
\newblock Konect: the koblenz network collection.
\newblock In {\em WWW}, pages 1343--1350. ACM, 2013.

\bibitem{lee19-motif-attention}
John~Boaz Lee, Ryan Rossi, Xiangnan Kong, Sungchul Kim, Eunyee Koh, and Anup
  Rao.
\newblock Graph convolutional networks with motif-based attention.
\newblock In {\em CIKM}, 2019.

\bibitem{VertexSimilarity}
E~A Leicht, Petter Holme, and M~E~J Newman.
\newblock Vertex similarity in networks.
\newblock {\em Physical review. E, Statistical, nonlinear, and soft matter
  physics}, 73 2 Pt 2:026120, 2006.

\bibitem{levy2014neural}
Omer Levy and Yoav Goldberg.
\newblock Neural word embedding as implicit matrix factorization.
\newblock In {\em NeurIPS}, 2014.

\bibitem{Lorrain}
Francois Lorrain and Harrison~C. White.
\newblock Structural equivalence of individuals in social networks.
\newblock {\em The Journal of Mathematical Sociology}, 1(1):49--80, 1971.

\bibitem{word2vec}
Tomas Mikolov, Ilya Sutskever, Kai Chen, Greg~S Corrado, and Jeff Dean.
\newblock Distributed representations of words and phrases and their
  compositionality.
\newblock In {\em NeurIPS}, 2013.

\bibitem{DD6_repo}
Christopher Morris, Nils~M Kriege, Franka Bause, Kristian Kersting, Petra
  Mutzel, and Marion Neumann.
\newblock Tudataset: A collection of benchmark datasets for learning with
  graphs.
\newblock In {\em ICML GRL Workshop}, 2020.

\bibitem{nandanwar2016structural}
Sharad Nandanwar and M~Narasimha Murty.
\newblock Structural neighborhood based classification of nodes in a network.
\newblock In {\em KDD}, pages 1085--1094, 2016.

\bibitem{segk}
Giannis Nikolentzos and Michalis Vazirgiannis.
\newblock Learning structural node representations using graph kernels.
\newblock {\em TKDE}, 2019.

\bibitem{ilprints422}
Lawrence Page, Sergey Brin, Rajeev Motwani, and Terry Winograd.
\newblock The pagerank citation ranking: Bringing order to the web.
\newblock Technical Report 1999-66, Stanford InfoLab, November 1999.
\newblock Previous number = SIDL-WP-1999-0120.

\bibitem{scikit-learn}
F.~Pedregosa, G.~Varoquaux, A.~Gramfort, and et~al.
\newblock Scikit-learn: Machine learning in {P}ython.
\newblock {\em JMLR}, 12:2825--2830, 2011.

\bibitem{deepwalk}
Bryan Perozzi, Rami Al-Rfou, and Steven Skiena.
\newblock Deepwalk: Online learning of social representations.
\newblock In {\em KDD}. ACM, 2014.

\bibitem{struc2vec}
Leonardo~FR Ribeiro, Pedro~HP Saverese, and Daniel~R Figueiredo.
\newblock struc2vec: Learning node representations from structural identity.
\newblock In {\em KDD}. ACM, 2017.

\bibitem{RossiA15}
Ryan~A. Rossi and Nesreen~K. Ahmed.
\newblock Role discovery in networks.
\newblock {\em TKDE}, 27(4):1112--1131, 2015.

\bibitem{rossi2019community}
Ryan~A Rossi, Di~Jin, Sungchul Kim, Nesreen~K Ahmed, Danai Koutra, and
  John~Boaz Lee.
\newblock From community to role-based graph embeddings.
\newblock {\em arXiv preprint arXiv:1908.08572}, 2019.

\bibitem{schnabel2015evaluation}
Tobias Schnabel, Igor Labutov, David Mimno, and Thorsten Joachims.
\newblock Evaluation methods for unsupervised word embeddings.
\newblock In {\em EMNLP}, pages 298--307, 2015.

\bibitem{wlgk}
Nino Shervashidze, Pascal Schweitzer, Erik Jan~van Leeuwen, Kurt Mehlhorn, and
  Karsten~M Borgwardt.
\newblock Weisfeiler-lehman graph kernels.
\newblock {\em Journal of Machine Learning Research}, 12(Sep):2539--2561, 2011.

\bibitem{line}
Jian Tang, Meng Qu, Mingzhe Wang, Ming Zhang, Jun Yan, and Qiaozhu Mei.
\newblock Line: Large-scale information network embedding.
\newblock In {\em WWW}, 2015.

\bibitem{drne}
Ke~Tu, Peng Cui, Xiao Wang, Philip~S. Yu, and Wenwu Zhu.
\newblock Deep recursive network embedding with regular equivalence.
\newblock In {\em KDD}, pages 2357--2366, 2018.

\bibitem{wasserman_faust_1994}
Stanley Wasserman and Katherine Faust.
\newblock {\em Social Network Analysis: Methods and Applications}.
\newblock Structural Analysis in the Social Sciences. Cambridge University
  Press, 1994.

\bibitem{gin}
Keyulu Xu, Weihua Hu, Jure Leskovec, and Stefanie Jegelka.
\newblock How powerful are graph neural networks?
\newblock In {\em ICLR}, 2019.

\bibitem{riwalk}
Ma~Xuewei, Geng Qin, Zhiyang Qiu, Mingxin Zheng, and Zhe Wang.
\newblock Riwalk: Fast structural node embedding via role identification.
\newblock In {\em ICDM}. IEEE, 2019.

\end{thebibliography}

\clearpage
\appendix
\section{Embedding Hyperparameters}
\label{app:embed-param}
Unless otherwise mentioned, our parameter settings for all methods follow default values suggested in the papers and/or official/available author implementations.  For convenience, here we cite the links to exact versions of the code and data we used for our experiments in addition to the citation of the papers that presented them in the body of our paper.  In order to make the comparison between the embedding methods fair, we transform all the input networks to be \textit{undirected} and \textit{unweighted}. For all methods, we learn 128-dimensional embeddings by default. 
\begin{itemize}
    \item For the skip-gram methods (node2vec, struc2vec~\citelatex{struc2vec_repo}, RiWalk~\citelatex{riwalk_repo}, and role2vec~\citelatex{role2vec_repo}), we sample context by performing 10 random walks (80 for struc2vec, which performs these walks on a more complex multi-layer structural similarity network) of length 80 per node.  We set the skip-gram window size to 10 and optimize the objective using 10 iterations of gradient descent. For scalability, we use all three optimizations for struc2vec and degree (or motifs, if applicable) as the feature for role2vec. 
    \item For node2vec~\citelatex{node2vec_repo}, we use random walk bias parameters $p = 1$ and $q = 4$ to tune node2vec to capture more structural equivalence, using parameter values considered in the original paper~\cite{node2vec}.
    \item For LINE~\citelatex{LINE_repo}, we set the \texttt{order} to be 2 and the total number of training samples to be 100 million and negative samples to be 5.
    \item For GraphWave~\citelatex{graphwave_repo}, we use the automatic selection method of the scale parameter~\cite{graphwave} and \texttt{exact} heat kernel matrix calculation.
    \item For struc2vec, xNetMF~\citelatex{xnetmf_repo}, and SEGK~\citelatex{segk_repo}, we consider up to 2-hop neighborhoods. In RiWalk, which also has a node neighborhood radius parameter $k$, we used default setting $k=4$. We discount the information of distant neighborhoods in xNetMF using a discount factor of 0.1 and set the structural similarity resolution parameter $\gamma = 1$.  For SEGK, we compare neighborhoods using the Weisfeiler-Lehman graph kernel~\cite{wlgk}.  We also use the Weisfeiler-Lehman graph kernel in RiWalk to identify structural roles of nodes based on their local neighborhoods (RiWalk-WL in \cite{riwalk}).
    \item For DRNE~\citelatex{DRNE_repo}, we follow the example usage to set the batch size to be 256 and the learning rate to be 0.0025.
    \item For MultiLENS~\citelatex{multilens_repo}, we set the \texttt{cat} input with all nodes having the same category/type. 
\end{itemize}

All sociological notions of equivalence are computed using the implementations of the CONCOR, MAXSIM, and CATREGE algorithms in the popular UCINET package~\citelatex{ucinet}. {The default settings in UCINET are adopted}. To compute vertex similarity (Section~\ref{subsec:real-data}), we use the implementation at \citelatex{VS_repo}.

\bibliographystylelatex{plain}
\bibliographylatex{BIB/bibliography}

\clearpage

\end{document}